\documentclass[reprint, showpacs, amsmath, amssymb, aip, rsi, superscriptaddress, floatfix]{revtex4-2}
\pdfoutput=1

\usepackage{graphicx}
\usepackage{xcolor}
\usepackage{dcolumn}
\usepackage{bm}
\usepackage{array}
\usepackage{siunitx}
\usepackage[capitalize]{cleveref}
\usepackage{multirow}
\usepackage{tabularx, booktabs}

\newcommand{\bafeas}{Ba(Fe$_{1-x}$Co$_x$)$_2$As$_2$}
\renewcommand\arraystretch{1.2}
\renewcommand{\vec}{\mathbf}
\graphicspath{{figures/}}

\renewcommand{\th}{\textsuperscript{th}}

\newcolumntype{Y}{>{\centering\arraybackslash}X}
\newcolumntype{C}[1]{>{\centering\let\newline\\\arraybackslash\hspace{0pt}}m{#1}}

\begin{document}

\preprint{APS/123-QED}

\title{Frequency-dependent sensitivity of AC elastocaloric effect measurements explored through analytical and numerical models}

\author{J. A. W. Straquadine}
\email{jstraq@stanford.edu}
\affiliation{Geballe Laboratory for Advanced Materials and Department of Applied Physics, Stanford University, California 94305, USA}
\affiliation{Stanford Institute for Materials and Energy Sciences, SLAC National Accelerator Laboratory, 2575 Sand Hill Road, Menlo Park, CA 94025, USA}

\author{M.S. Ikeda}
\affiliation{Geballe Laboratory for Advanced Materials and Department of Applied Physics, Stanford University, California 94305, USA}
\affiliation{Stanford Institute for Materials and Energy Sciences, SLAC National Accelerator Laboratory, 2575 Sand Hill Road, Menlo Park, CA 94025, USA}

\author{I. R. Fisher}
\affiliation{Geballe Laboratory for Advanced Materials and Department of Applied Physics, Stanford University, California 94305, USA}
\affiliation{Stanford Institute for Materials and Energy Sciences, SLAC National Accelerator Laboratory, 2575 Sand Hill Road, Menlo Park, CA 94025, USA}

\date{\today}

\begin{abstract}

	We present a comprehensive study of the frequency-dependent sensitivity for measurements of the AC elastocaloric effect by applying both exactly soluble models and numerical methods to the oscillating heat flow problem.
	These models reproduce the finer details of the thermal transfer functions observed in experiments, considering here representative data for single-crystal \bafeas{}.
	Based on our results, we propose a set of practical guidelines for experimentalists using this technique.
	This work establishes a baseline against which the frequency response of the AC elastocaloric technique can be compared and provides intuitive explanations of the detailed structure observed in experiments.
\end{abstract}

\maketitle

\section{Introduction}

\begin{figure}
	\centering
	\includegraphics[width=\columnwidth]{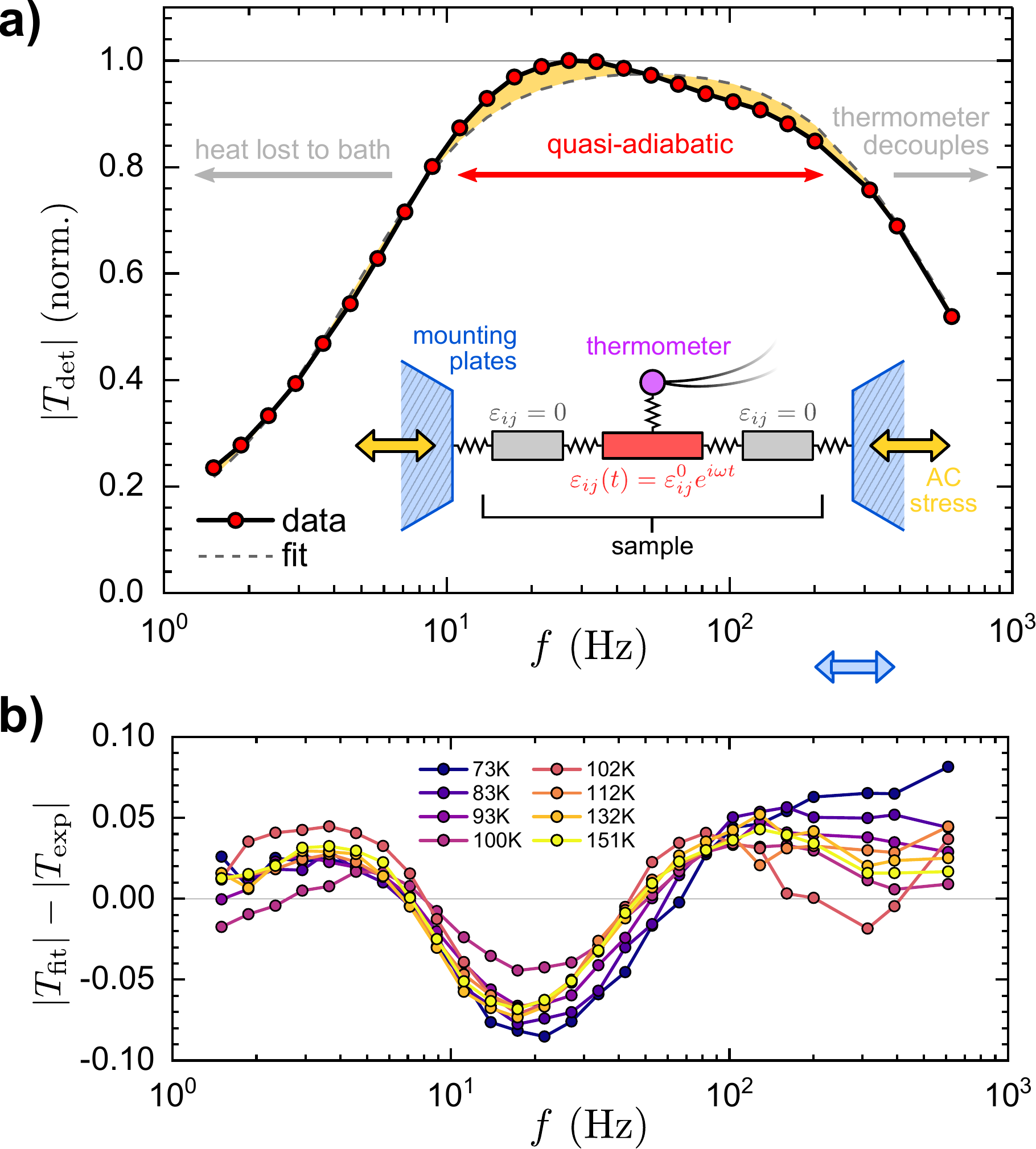}
	\caption[Frequency dependent magnitude of the temperature dependent oscillation signal observed in the AC-ECE.]{The frequency-dependent magnitude of the temperature oscillation signal observed in the AC-ECE.
		(a) Magnitude of the observed temperature oscillation $T_{det}$ as a function of frequency for an AC-ECE measurement of a sample of \bafeas{} at 100~K.
		Shaded regions highlight the deviations of the data from the best fit prediction for the simplest discrete model.
		A schematic of this model is shown in the inset.
		(b) Residuals of fits like the one shown in panel (a) for a series temperatures.
		The primary goal of this paper is to provide an understanding of the consistent deviations in experimental data from the fits.
	}
	\label{fig:intro}
\end{figure}

	The elastocaloric effect (ECE) describes the dependence of a material's entropy on externally imposed strain.
	This can be quantified either by measuring the change in entropy $S$ resulting from  isothermal changes in strain $\varepsilon_{ij}$, or by measuring the change in temperature resulting from adiabatic changes in strain.
	For the purposes of this work, we adopt the latter definition.
	The simplest and most common technique for measuring the ECE is simply to measure the temperature during a single rapid application of compressive or tensile strain.
	This technique has been applied extensively in studying materials such as shape-memory alloys, which show significant promise for solid-state elastocaloric refrigeration due to the high latent heat of strain-induced martensitic transitions.\cite{Ziotkowski1993,Cui2012,Qian2016,Qian2016a,Luo2017}
	The ECE is, however, a much more general feature of the solid state; any change of entropy induced by strain, regardless of the microscopic details, is necessarily reflected in an elastocaloric temperature change under adiabatic conditions.
	While few materials are expected to have as dramatic an ECE response as shape-memory alloys, uniaxial stress and the accompanying symmetry-breaking strains have emerged in recent years as versatile tuning parameters for the phases and phase transitions in several families of strongly correlated materials.\cite{Chu2012,Hicks2014a,Kim2018a,Bachmann2019,Rosenberg2019}
	Consequently, ECE measurements have the potential to directly probe changes in the entropy of these strain-sensitive materials.
	However, the large deformations required for conventional ECE measurements are not well-suited to these materials, which are often brittle, highly anisotropic, or cleave easily under stress.
	With this in mind, Ikeda et al \cite{Ikeda2019} developed the use of small oscillating (AC) strains to measure the elastocaloric effect.
	In an AC-ECE measurement, a bar-shaped sample is glued or clamped between a pair of mounting plates, which then apply a small oscillating stress to the sample, as shown in \cref{fig:intro}(a).
	The resulting temperature oscillations are detected with a small thermometer (usually a thermocouple) attached to the center of the freestanding section.

	The development of the AC-ECE technique presents several benefits for fundamental research.
	Firstly, AC-ECE enables the use of phase-sensitive detection, enabling long averaging times and high resolving power.
	Furthermore, the ability to detect small signals makes it possible to use much smaller strains than conventional techniques, preventing sample fatigue effects and operating in a regime of linear response to changes in both strain and temperature.
	Modern piezoelectric technology easily facilitates the \textit{in situ} application of small oscillating stresses at temperatures spanning from above room temperature to below 1 K, such that a single apparatus can access large regimes of phase space.\cite{Hicks2014}
	Finally, by introducing frequency as a new tuning parameter, the AC-ECE creates the possibility of exploring dynamical effects intrinsic to the sample material, such as the motion of domain walls\cite{Hristov2019a}.

	At the most basic level, the elastocaloric effect can be modeled as a series of discrete thermal elements, as illustrated in \cref{fig:intro}(a).
	A sample with heat capacity $C_s$ is coupled with thermal conductance $K_b$ to a thermal reservoir at temperature $T_0$ and with thermal conductance $K_t$ to a thermometer of heat capacity $C_t$.
	We define thermal relaxation times of the sample and thermometer as $\tau_s=C_s/K_b$ and $\tau_t=C_t/K_t$, respectively, and assume that $\tau_t<\tau_s$.
	The sample is then exposed to endogenous heating and cooling within the strained section which oscillates sinusoidally at frequency $f$.

	This simple model describes the asymptotic behavior of the AC-ECE.
	In the limit $f \ll \tau_s^{-1}$, the temperature of the thermometer $T_t$ and the sample $T_s$ only experience small oscillations about the bath temperature $T_0$.
	In the limit of $f \gg \tau_t^{-1}$, the sample temperature oscillates around $T_0$ with amplitude $T_\infty=QC_s^{-1}$, where $Q$ represents the effective heat generated by the elastocaloric effect.
	The thermometer temperature, however, again performs vanishingly small oscillations around $T_0$ due to the finite thermal relaxation time of the thermometer.
	The magnitude of the thermometer temperature oscillation reaches a maximum at intermediate frequencies, and the transfer function has a flat plateau in the range $\tau_s^{-1} < f < \tau_t^{-1}$ with no dependence on frequency.

	As a concrete example, we consider the case of a prototypical iron-based superconductor material, \bafeas{}.
	These materials exhibit a coupled electronic nematic/orthorhombic phase transition as well an antiferromagnetic transition, both of which have been shown to be highly sensitive to strain.\cite{Ikeda2018}.
	The frequency-dependent AC-ECE signal is presented in \cref{fig:intro}(a) for a sample of this material for $x=0.021$ measured at 100~K.

	The best fit of the simple discrete model to the experimental result is plotted on the same axis, and it can easily be seen that the experimental results deviate in several nontrivial ways.
	First, the corner between the low and intermediate frequency ranges always appears significantly sharper than this model would predict.
	Secondly, the predicted flat plateau is replaced by a sloping shoulder.
	Incorporating a frequency dependence to the ratio of the thermometer and sample heat capacities (a consequence of a finite thermal length $\xi\equiv (D/f)^{-1/2}$, where $D=k_s/C_s$ is the thermal diffusivity of the sample\cite{Riou2004,Ikeda2019}) does slightly suppress the high frequency response, but improvement in fit quality is minor.

	The consistent behavior of the residuals of the fit (\cref{fig:intro}(b)) indicates that this simple model overlooks some nontrivial details in the frequency-dependent sensitivity of the AC-ECE technique
	Without a quantitative theoretical understanding of the frequency-dependent behavior of the AC-ECE, the overall signal magnitude cannot be ascertained with confidence.
	Also, the present understanding of the details of heat transfer during a measurement is also insufficient to identify or rule out experimental artifacts which could contribute to these deviations.
	Finally, frequency-dependent dynamical behavior intrinsic to the sample will also be at least partially masked by the experimental sensitivity.
	A thorough understanding of the practical frequency dependence effects is critical to interpreting the empirical results and in benchmarking the quality of a given measurement.
	This work seeks to establish both a detailed description and an intuitive interpretation of the frequency-dependent sensitivity inherent in thermocouple-based AC-ECE measurements.

	We begin in Section \ref{sec:formalism} by establishing a formal definition of the frequency-dependent AC-ECE sensitivity function $\Gamma(\omega)$.
	In Section \ref{sec:toymodels}, we present a pair of exactly soluble models for the low-frequency component of $\Gamma(\omega)$ which exhibit and provide an intuitive basis for understanding the low frequency behavior.
	We then describe the setup and implementation for our finite element calculations in Section \ref{sec:model}, and then use this model to describe the contributions to $\Gamma(\omega)$ arising from practical effects such as sample mounting and thermometer characteristics in Section \ref{sec:results}.
	Finally in Section \ref{sec:data} we solve for the full sensitivity function with the finite element method, revisit the comparison of these results to experimental results, and provide an empirical method for estimating the peak sensitivity in a given measurement.

\section{Statement of the problem}\label{sec:formalism}
	Consider a bar-shaped sample composed of a material with volumetric heat capacity $c(\vec{r})$ and thermal conductivity $k(\vec{r})$.
	We define the elastocaloric tensor $\eta_{ij}$ of the sample material as
	\begin{equation}\label{eq:defeta}
		\eta_{ij} \equiv \left( \frac{dT}{d\varepsilon_{ij}}\right)_S
	\end{equation}
	and we assume for now that the sample experiences a spatially homogeneous time-varying strain $\varepsilon_{ij} = \varepsilon_{ij}^0 e^{i\omega t}$.
	In general, $\eta_{ij}$ may carry a frequency dependence and may take on complex values $\eta_{ij} = \eta_{ij}^{\prime}(\omega) + i\eta_{ij}^{\prime\prime}(\omega)$ reflecting the dynamical behavior of the material.
	However, for the purposes of this work we enforce $\eta_{ij}^{\prime\prime}(\omega)=0$ and $d\eta_{ij}/d\omega=0$.
	The sample temperature oscillation in the adiabatic limit $T_\infty$ will therefore be given by
	\begin{equation}\label{eq:deftinf}
		T_\infty(\vec{r},t) = \eta_{ij}\varepsilon_{ij}^0(\vec{r}) e^{i\omega t}.
	\end{equation}
	The goal of the AC-ECE technique is to use this oscillating temperature signal to accurately quantify $\eta_{ij}$.

	Deviations from the ideal adiabatic limit due to practical constraints can be described by a complex-valued sensitivity function $\Gamma(\omega)$ such that the temperature oscillation $T_\mathrm{det}(t)$ detected by a thermometer at position $\vec{r_0}$ is given by
	\begin{equation}\label{eq:defgamma}
		T_\mathrm{det}(t) = \Gamma(\omega)T_\infty(t) = \Gamma(\omega)\eta_{ij}\varepsilon_{ij}^0 e^{i\omega t}
	\end{equation}
	We can separate the impacts of different practical limitations by writing the sensitivity as a product ${\Gamma(\omega) = \Gamma_b(\omega)\Gamma_t(\omega)}$.
	Here $\Gamma_b(\omega)$ describes the loss of heat due to coupling to the bath, which is the dominant deleterious effect at low frequencies.
	$\Gamma_t(\omega)$ describes sensitivity losses due to poor coupling of the thermometer to the sample, which dominates at high frequencies.

	The sound velocity in most solids is typically several thousand meters per second; for a millimeter-scale sample, a sound wave traverses the entire sample in $\lesssim$~\SI{1}{\micro s}.
	Current technology imposes an upper bound on the range of accessible strain frequencies at approximately 10~kHz.
		\footnote{Piezoelectric actuators can, and often are, operated at strain frequencies well into the ultrasound range.
		However, incorporating piezoelectric actuators into a uniaxial stress cell assembly\cite{Hicks2014} lowers resonance frequencies and creates a risk of vibrational fatigue within epoxies and the piezoelectric actuators themselves.
		Careful design of future devices may raise this practical upper bound.}
	The minimum applicable strain oscillation period is then two orders of magnitude greater than both the shock propagation time as well as the estimated thermoelastic relaxation times for most solids.\cite{Hetnarski2009}
	Additionally, the amplitude of both the strain and temperature oscillations are assumed to be small, which justifies the use of constant values of $c(\vec{r})$ and $k(\vec{r})$.
	(The position dependence of these parameters reflects only the possibility of regions of different materials, such as the sample and the mounting plates--the heat capacity and thermal conductivity are assumed to be homogeneous throughout a given material.)
	As a consequence, the simplest Fourier heat flow model can be expected to capture the observable phenomena without requiring recourse to a full set of hyperbolic thermoelastic partial differential equations.\cite{Biot1956,Lord1967,Green1972}
	Similarly, second sound effects are not included in this work.

	Linear response (small strain oscillations and small temperature oscillations) justifies the use of the standard heat equation in describing heat transfer in AC-ECE measurements:
	\begin{equation}\label{eq:heatequation}
		c(\vec{r})\frac{dT(\vec{r}, t)}{dt} = k(\vec{r})\nabla^2 T(\vec{r}, t) + Q(\vec{r},t)
	\end{equation}
	where $T(\vec{r},t)$ is the temperature profile within the sample, and the source term $Q(\vec{r},t)$ simulates the elastocaloric effect.
	This source term does not represent conductive, radiative, or convective heat transfer between the sample and another body, but rather the redistribution of entropy between various microscopic subsystems within the sample.
	One example would be entropy due to fluctuations in electronic degrees of freedom near a continuous phase transition--if a change in the sample strain alters the total entropy in these fluctuations, isentropic conditions dictate that the sum of all other degrees of freedom in the sample (phonons, magnetic moments, etc.) must experience an equal and opposite change in entropy, which is reflected in a change in temperature.
	Away from the adiabatic limit heat may flow into or out of the sample, and this will suppress the observed temperature oscillation.
	However, this effect is described solely by the Fourier heat conduction term proportional to $k(\vec{r})$, not the elastocaloric heat generation term $Q(\vec{r})$.
	\footnote{
		Changes of strain imply that work is being done on the sample, changing the internal elastic energy even in the adiabatic limit.
		This would also result in a change in the sample temperature; but since both compressive and tensile deformations of a solid at equilibrium requires positive work, an oscillation at frequency $f$ about the strain-neutral condition generates temperature oscillations at frequency $2f$, which will not affect phase sensitive measurements.
		If the oscillation is superimposed on a constant strain offset, there \emph{will} be a contribution to the signal at $f$ arising from the work done on the sample.
		Both the conventional elastocaloric effect and the elastic energy away from the strain-neutral point, however, can be lumped together into our definition of $Q(\vec{r},t)$.
	}

	With these physical definitions, \cref{eq:heatequation} holds regardless of the functional form of $Q(\vec{r},t)$.
	However, we specialize to the case of sinusoidal strains, taking as an ansatz $Q(\vec{r},t)=Q_0e^{i\omega t}$.
	We define the magnitude and phase $Q_0$ of the heat term in the adiabatic limit $\omega\rightarrow\infty$, or, equivalently, $k\rightarrow 0$.
	In this limit, \cref{eq:defgamma,eq:heatequation} show that
	\begin{equation}
		Q_0(\vec{r}) = i\omega\eta_{ij}c\varepsilon_{ij}^0(\vec{r}).
	\end{equation}
	Equivalently, the source term in \cref{eq:defgamma} is linear in the ECE tensor, the volumetric heat capacity, and the rate of change of strain.
	Throughout this work, we assume that the magnitude of the strain oscillation is known--this must be measured independently of the temperature oscillation.
	Techniques for quantifying oscillating strain are beyond the scope of this paper, and the reader is referred to refs. \onlinecite{Hristov2018,Straquadine2019a,Straquadine2020} for three different options.

	All that remains is to define a suitable geometry and boundary conditions under which to solve \cref{eq:heatequation} for $T_\mathrm{det}(t)=T(\vec{r_0},t)$, which, given $\eta_{ij}$ and $\varepsilon_{ij}^0$, allows the characterization of $\Gamma(\omega)$.
	We begin by examining two models which can be solved exactly and which isolate $\Gamma_b(\omega)$.
	We then explore both $\Gamma_t(\omega)$ and the full $\Gamma(\omega)$ numerically through the finite element method.

\section{Exactly soluble models for $\Gamma_b(\omega)$}\label{sec:toymodels}
	\subsection{Continuum model}\label{sec:continuum}
		\begin{figure}
			\centering
			\includegraphics[width=\columnwidth]{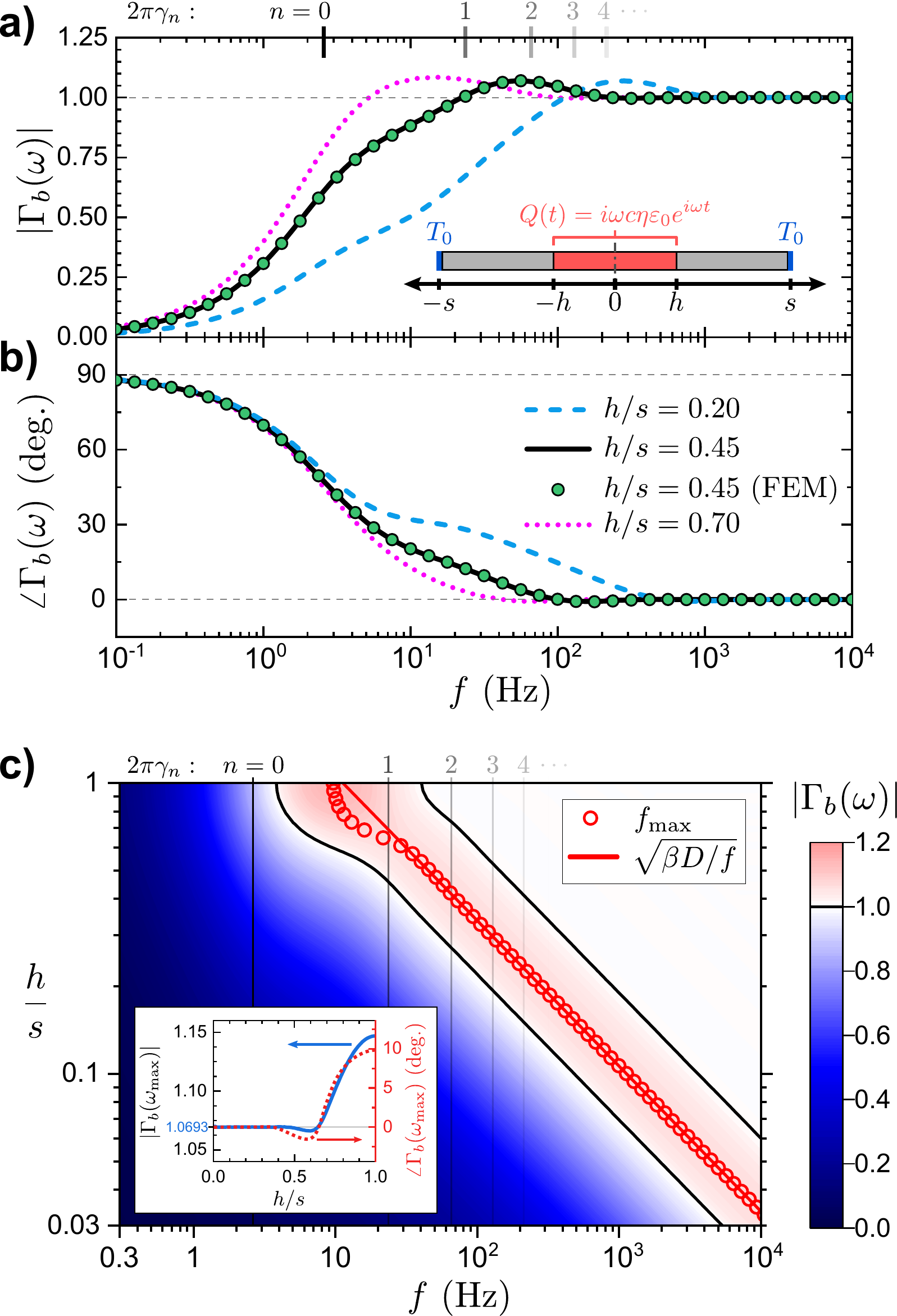}
			\caption[Low frequency component $\Gamma_b(\omega)$ of the AC-ECE sensitivity calculated using the 1D continuum model.]{Low frequency component $\Gamma_b(\omega)$ of the AC-ECE sensitivity calculated using the 1D continuum model.
				Magnitude (a) and phase (b) of the sensitivity function as a function of frequency for several different ratios of sample length $s$ to the length of the thermally excited region $h$.
				Lines correspond to the result of evaluating \cref{eq:contgamma}, and filled circles denote FEM solutions for a three-dimensional bar-shaped sample with the same geometry and boundary conditions along one axis.
				All calculations use the same values of the sample specific heat and thermal conductivity, as described in Section \ref{sec:data}.
				A schematic of the geometry and boundary conditions is shown in the inset of panel (a).
				The $h/s$ ratio alters the location of the maximum in $|\Gamma_b(\omega)|$, and affects the slope of the curve for slightly lower frequencies, but otherwise preserves the general structure.
				Smaller values of $h/s$ retain a finite phase lag out to higher frequencies.
				Magnitude values are plotted for a larger parameter space in panel (c), which allows the structure and geometry dependence to be seen more clearly.
				Dark lines enclose the region of $|\Gamma_b(\omega)|>1$, which exists for any geometry, but which moves to higher $f$ for smaller $h/s$.
				Vertical lines indicate the characteristic relaxation rates of the thermal modes.
				Inset: magnitude and phase of $\Gamma_b(\omega)$ at the peak frequency as a function of $h/s$.
			}
			\label{fig:continuum}
		\end{figure}

		We begin with a simple one-dimensional continuum model for heat flow within the sample.
		We discard the geometry of the thermometer and mounting plates, and consider the sample as a one-dimensional object for ${x\in[-s,s]}$, as shown in \cref{fig:continuum}(a).
		Neglecting the thermometer allows us to set $\Gamma_t(\omega)=1$.
		We denote by $c$, $k$ and $T(x,t)$ the heat capacity per unit length, the thermal conductance, and the temperature distribution within the sample.
		At either end of the sample, we dictate that $T(x=\pm s,t) = T_0$
		We assume that the region ${x\in[-h,h]}$ is subjected to homogeneous strain $\varepsilon=\varepsilon_0 e^{i\omega t}$, resulting in
			\begin{equation}\label{eq:contq}
					Q(x,t) = Q_0 \left[ \Theta(x+h) - \Theta(x-h)\right] e^{i \omega t}
			\end{equation}
		where $\Theta(x)$ is the Heaviside step function, $Q_0 = i\omega\eta c\varepsilon_0$, and where we have dropped the tensor indices on $\eta$ and $\varepsilon$ for brevity.

		The solution for $T(x,t)$ can be determined through a straightforward eigenfunction expansion
		\begin{equation}\label{eq:eigexp}
			T(x,t) = \sum_{n=0}^\infty A_n(t)B_n(x)
		\end{equation}
		where the $n$\th{} spatial mode is described by $B_n(x) = \cos{(\xi_n^{-1} x)}$ and where
		\begin{equation}
			\xi_n  = \frac{2s}{\pi(2n+1)}
		\end{equation}
		is the characteristic thermal length of the $n$\th{} mode.
		The steady-state amplitude of the $n$\th{} mode can be evaluated to be
		\begin{equation}
			A_n(t) = \frac{4 \eta \varepsilon_0}{\pi(2n+1)} \sin(\xi_n^{-1} h) \left(\frac{\omega}{\omega - i\gamma_n}\right)e^{i \omega t}
		\end{equation}
		where $\gamma_n = \xi_n^{-2}D$ is the characteristic thermal relaxation and $D=k/c$ is the thermal diffusivity.

		Now suppose an ideal thermometer ($\Gamma_t(\omega)=1$) is placed at $x=0$.
		The detected temperature $T_\mathrm{det}(t) = T(x=0,t)$ is given by
		\begin{equation}
			T_\mathrm{det}(t) = \sum_n A_n(t) = \Gamma_b(\omega)\eta\varepsilon_0 e^{i\omega t}
		\end{equation}
		resulting in
		\begin{equation}\label{eq:contgamma}
			\Gamma_b(\omega)  = \frac{4}{\pi}\sum_n \left[ \frac{\sin(\xi_n^{-1} h)}{(2n+1)}  \left(\frac{\omega}{\omega - i\gamma_n}\right) \right]
		\end{equation}

		The magnitude and phase of $\Gamma_b (\omega)$ for realistic values of $k$, $c$, $h$, and $s$ is shown in \cref{fig:continuum}.
		As expected, $|\Gamma_b(\omega)|$ approaches zero as $\omega\rightarrow 0$ and approaches unity as $\omega\rightarrow\infty$.
		At an intermediate frequency which we label as $f_\mathrm{max}=2\pi\omega_\mathrm{max}$, however, we observe a global maximum in $|\Gamma_b(\omega)|$ where the sensitivity function reaches values greater than unity.
		\Cref{fig:continuum}(c) shows $|\Gamma_b(\omega)|$ for a range of values of $h/s$, demonstrating that the presence of such a peak is not an artifact of the geometry.
		Holding the thermal parameters constant, decreasing the proportion of the sample which is heated pushes the maximum out to higher frequencies.
		For $h/s \lesssim 0.6$, when $f_\mathrm{max}\gtrsim 2\pi\gamma_1$, the peak frequency is closely approximated by $f_\mathrm{max}\approx (h/s)^2(\beta D)^{-1}$, where $\beta$ is a constant with value $\approx 1.687$ and $D$ is the thermal diffusivity.
		The peak magnitude is largest for $h/s=1$, at which $|\Gamma_b(\omega_\mathrm{max})|=1.147$, and for $h/s\lesssim 0.5$ the amplitude approaches a constant value of 1.0693.
		Put another way, this peak always occurs as the characteristic diffusion length becomes smaller than the length of the excited region.
		These calculations suggest that this peak is a robust feature of the AC-ECE sensitivity function.
		Physically, the temperature gradient arising from the spatial variation of the phase of temperature oscillations generates a retarded flow of heat (and therefore entropy) within the sample.
		The phase delay between the elastocaloric heating and the conductive heat flow can boost the total rate of change of the temperature at and near the center of the sample.
		The results presented in \cref{fig:continuum} show that decreasing the thermal conductivity does not remove this peak, but rather pushes the peak to higher frequencies.
		We also refer the reader to Section \ref{sec:discrete} for an equivalent understanding of this peak motivated from the principle of superposition.

		Additionally, the thermal parameters $k$ and $c$ only enter \cref{eq:contgamma} in the form of the diffusivity ratio $D=k/c$ within the definition of $\gamma_n$; scaling $D$ by a multiplicative factor simply rescales the frequency axes in \cref{fig:continuum} by the same factor.

		As a verification of these results as well as a benchmark for the numerical methods used in Section \ref{sec:results} used here, we have also performed finite element simulations for similar conditions.
		We used a three-dimensional rectangular prism of length $2s$, width $w$, and thickness $d$, with specific heat $c_\mathrm{3D} = (wd)^{-1}c$ and thermal conductivity $k_\mathrm{3D}=(wd)^{-1}k$, with the same heating and boundary conditions.
		The results obtained through both numerical and analytical methods match quantitatively.

		In practice, the sample mounting plates generate more complicated boundary conditions; the epoxy holding the sample in place will allow for finite heat flow for $h<x<s$.
		However, as the freestanding region $-h<x<h$ is still thermally isolated on all other surfaces, this nonideality will only result in a smaller effective value of $s$.
		This will tend to increase $h/s$, resulting in a lower peak frequency for the same $h$, $k$, and $c$.
		This acts in the experimentalist's favor by increasing the frequency range over which $\Gamma(\omega)| \sim 1$.
		This effect is quantified in FEM calculations presented in Section \ref{sec:bath}.

	\subsection{Discretized model}\label{sec:discrete}
		\begin{figure}
			\centering
			\includegraphics[width=\columnwidth]{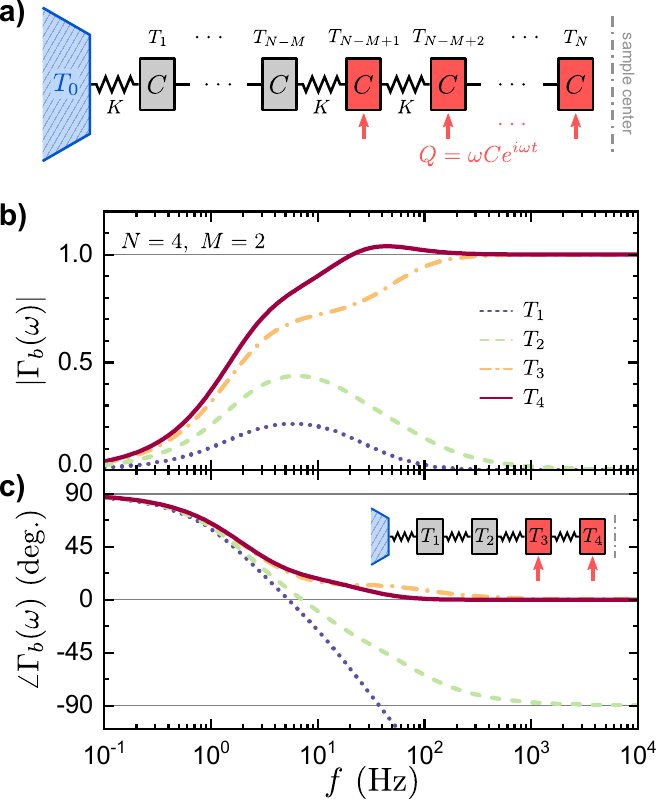}
			\caption[Discretized model of heat flow in the AC-ECE.]{Discretized model of heat flow in the AC-ECE.
				(a) Schematic of the model, in which a 1D sample is collected into $2N$ identical discrete elements.
				Only one half needs to be considered due to symmetry about the sample center.
				The $M$ units closest to the middle are subjected to an oscillating heat term.
				(b) Magnitude and (c) phase of the low frequency sensitivity function $\Gamma_b(\omega)$ for $N=4$ and $M=2$.
				We have set both $\eta$ and $\varepsilon$ to unity such that the observed temperature oscillation $T_n$ is identical to the sensitivity function for a thermometer placed on the $n$\th{} element.
				Similar to \cref{fig:continuum}, a peak greater than unity is observed in $T_4$ at intermediate frequencies.
			}
			\label{fig:discrete}
		\end{figure}

		The results of the previous section, are not limited to continuous thermal models.
		We now examine a minimal lumped-element thermal model which reproduces the same behavior and provides further intuition for the physical meaning of $|\Gamma_b(\omega)|>1$.
		We begin by separating the 1D sample of the previous section into a linear chain of $2N$ discrete elements ($N\ge 2$), each with heat capacity $C$ and coupled to its nearest neighbors with thermal conductance $K$.
		We maintain the same thermal behavior by defining $C= c(s/N)$ and $K=k(s/N)$.
		The first and last elements are also coupled to a heat bath at temperature $T_0$ with thermal conductance $K$.
		By merit of the symmetry of the chain, we safely neglect one half and only consider the first $N$ elements as shown in \cref{fig:discrete}(a).

		We denote the temperature of the $i$\textsuperscript{th} element by $T_i(t)$, and collect these terms into a vector $\vec{T} = (T_1~T_2~\cdots~T_N)^T$.
		The $M$ elements closest to the center of the sample ($M\le N$) are subjected to an oscillating heat term
		\begin{equation}
		Q_i(t) =
		\begin{cases}
		Q_0 e^{i\omega t}		&\text{where}~(N-M) < i\le N \\
		0										&\text{otherwise}
		\end{cases}
		\end{equation}
		The equations for heat flow between elements, taking as our ansatz $T_i = T_0 + T_i^0 e^{i\omega t}$, can be collected into matrix form $\vec{Q} = A\vec{T}$ where $A$ is a tridiagonal matrix given by
		\begin{equation}
		\renewcommand\arraystretch{1.5}
		A = \begin{pmatrix}
			g & K &        &        &       \\
			K & g & K      &        &       \\
			  & K & \ddots & \ddots &       \\
			  &   & \ddots & g      & K     \\
			  &   &        & K      & (g+K)
		\end{pmatrix}
		\end{equation}
		where ${g = i\omega C - 2K}$, and all omitted elements vanish.
		The extra unit of $K$ in element $A_{NN}$ reflects the fact that no heat flows across the mirror plane in the sample.
		The temperature profile can be computed immediately as ${\vec{T} = A^{-1}\vec{Q}}$.
		We finally extract the low-frequency sensitivity function $\Gamma_b$ by setting the elastocaloric tensor $\eta$ and strain magnitude $\varepsilon_0$ to unity, which by \cref{eq:defgamma} equates the magnitude and phase of $T_i$ with that of $\Gamma_b$ for an ideal thermometer placed on the $i$\th{} element.
		For all cases for which of $N\ge M\ge 2$, the elements $i$ for which $N-M+2\le i$ exhibit a peak temperature oscillation magnitude greater than unity.

		The principle of superposition, afforded by the linearity of \cref{eq:heatequation}, provides an intuitive explanation for the appearance of this peak.
		Consider first a case where $N=2$ and $M=1$; this model is related to the model first applied to interpreting AC-ECE measurements, but with the assumption of an ideal thermometer placed at $i=2$.
		In this case, $|\Gamma_b(\omega)|$ will increase monotonically with frequency and will not generate a peak.
		In contrast, consider a case in which $N=2$ but heat is applied only to the $i=1$ element, not the $i=2$ element, which is reminiscent of the model used in AC heat capacity measurements.\cite{Sullivan1968,Velichkov1992}.
		Here $|\Gamma_b(\omega)|$ (still as measured at $i=2$) will vanish at either frequency limit, but will have a finite peak below unity at some intermediate frequency.

		The case of $N=M=2$, the simplest case in which $|\Gamma_b(\omega)|>1$, can be considered the superposition of these two cases.
		If the peak in the second case occurs at or above the frequency at which the oscillation amplitude in the first case approaches unity, then the total response can result in a peak amplitude greater than unity.
		This is always the case for the model considered here due to the equal heat capacity of and thermal conductances between the elements.
		In summary, the peak in $|\Gamma_b(\omega)|$ is a robust feature which contributes to the thermal transfer function of all AC-ECE measurements.
		It is possible, however, that the peak itself is masked by the high-frequency component of the sensitivity function, $\Gamma_t(\omega)$, which will be discussed in later sections.

\section{FEM Implementation}\label{sec:model}

	\begin{figure*}
		\includegraphics[width=\textwidth]{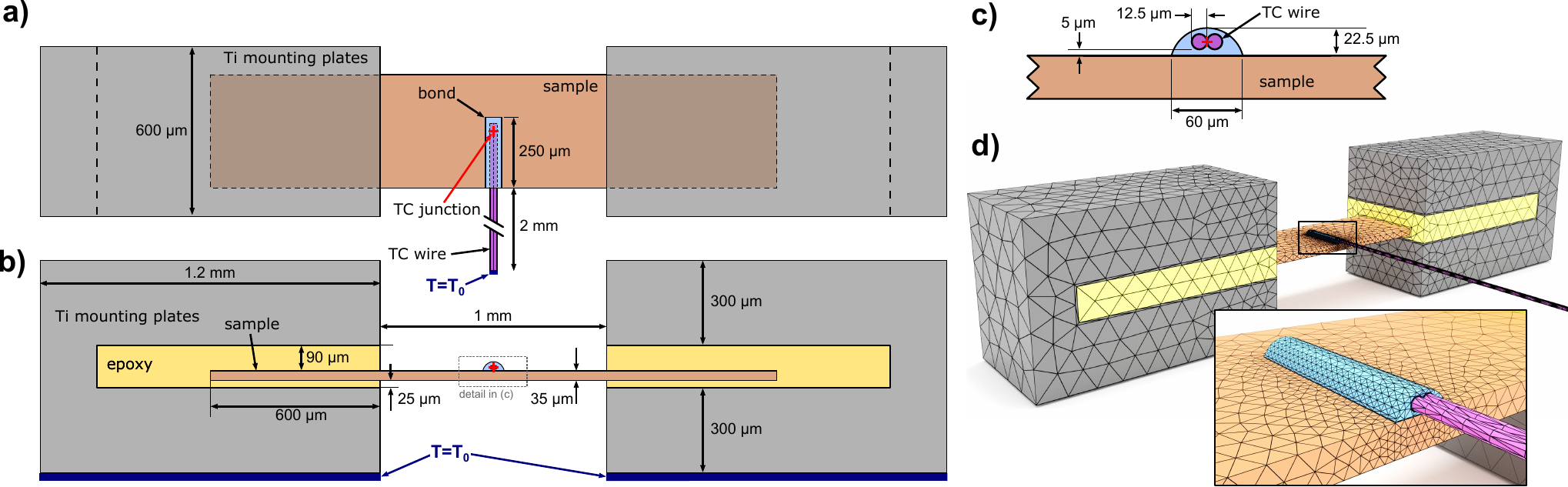}
		\caption[Geometry of the experimental setup, seen in top (a) and side (b) view.]{Geometry of the experimental setup, seen in top (a) and side (b) view.
				(c) Detailed side view of the thermocouple bond bead and thermocouple wire.
				(d) Rendered image of the full mesh and the thermocouple bond.
				Characteristic element sizes near the center of the sample are \SI{10}{\micro m}, expanding to \SI{75}{\micro m} at the edge of the mounting plates.
				As several variations of this standard mesh are used throughout this work, differences from this geometry are described in the text.
			}
		\label{fig:ecgeo}
	\end{figure*}

	We now turn our attention to nontrivial effects of the 3D geometry of the sample, mounting plates, and thermometer.
	We apply the finite element method (FEM) to solve \cref{eq:heatequation} on a 3D mesh in which regions representing different materials are assigned appropriate thermal conductivities and heat capacities.
	We use a Python implementation of the open-source FEniCS Project\cite{Alnaes2015,Logg2010} as our FEM solver.
	Generation and optimization of the mesh was performed using Gmsh.\cite{Remacle2012}

	The complete finite element mesh, slightly simplified from practical experiments, is shown in \cref{fig:ecgeo}.
	Each pair of mounting plates, as well as the screws which hold the assembly together, are represented by a single C-shaped block.
	The regions between the mounting plates are completely filled with epoxy, into which either end of the sample is then embedded.
	The thermometer used to detect the temperature oscillation is taken to be a thermocouple (TC) and is represented by two cylinders approximately 2~mm in length.
	One end is ``adhered'' to the sample with a bead of some bonding material.
	In practice such material could be an epoxy, an electrically conductive paint, or a small quantity of solder.
	For simplicity, the TC wires are taken to lie parallel and adjacent to each other and the TC ``junction'' is defined as the point directly between the wires and above the center of the sample.
	We describe the glue bead as a section of a cylinder \SI{60}{\micro m} in diameter and \SI{250}{\micro m} in length, which spans from one edge of the sample to slightly beyond the end of the TC wire.
	Several studies in Section \ref{sec:results} employ meshes which remove either the mounting plates or the TC in order to isolate the $\Gamma_b(\omega)$ and $\Gamma_t(\omega)$ components independently.

	Characteristic dimensions of the tetrahedral mesh elements within the sample vary from less than \SI{10}{\micro m} near the center and within the TC to approximately \SI{75}{\micro m} where the sample meets the mounting plates.
	Elements composing the mounting epoxy and the mounting plates increase from \SI{75}{\micro m} to \SI{125}{\micro m} at the outside edge.
	As will be shown in later sections, the temperature variation in these regions is negligibly small for realistic parameters, which makes additional refinement of the mesh in and around the mounting plates unnecessary.

	\begin{table}[]
		\caption[Baseline thermal parameters used for the calculations of the AC-ECE.]{Baseline thermal parameters used for the calculations in this paper.
			We emulate an AC-ECE measurement on a sample of \bafeas{} at a temperature of 100~K.
			The thermocouple is assumed to be Type E (chromel-constantan\cite{Croarkin1993}).
		}\label{tab:params}
	\begin{tabularx}{\columnwidth}{rYYY}
		\hline
		& \multicolumn{1}{c}{\begin{tabular}[c]{@{}c@{}}k\\ (Wm$^{-1}$K$^{-1}$)\end{tabular}} & \multicolumn{1}{c}{\begin{tabular}[c]{@{}c@{}}C\\ (J cm$^{-3}$K$^{-1}$)\end{tabular}} & \multicolumn{1}{c}{Refs.}              \\ \hline
		sample       & \multicolumn{1}{c}{\begin{tabular}[c]{@{}c@{}} $k_{xx}=10.6$\\ $k_{zz}=3.2$\end{tabular}}                                                                              & 1.31                                                                                  & [\onlinecite{Machida2009a,Chu2009,Meinero2019}]             \\
		titanium     & 9.6                                                                                 & 1.36                                                                                  & [\onlinecite{Corruccini1960,Schwartzberg1970}] \\
		thermocouple & 15.6                                                                                & 2.11                                                                                  & [\onlinecite{Sundqvist1992}]                   \\
		sample/TC epoxy & 0.19                                                                                & 0.497                                                                                 & [\onlinecite{Nakamura2018}]                    \\ \hline
	\end{tabularx}
\end{table}

	The section of the sample which is suspended between the plates experiences a fairly homogeneous strain environment.
	However, finite but inhomogeneous strain persists within the glued ends of the sample.
	This strain will still contribute to the overall elastocaloric response of the sample.
	Using the strain profile reported in ref. \onlinecite{Ikeda2018} for a similar setup, we find that the strain within the glued sections can be approximated by
	\begin{equation}
	\varepsilon_{xx}(x) = 0.0286 \exp\left(\frac{-(x+597)}{168}\right)
	\end{equation}
	where $x$ is the depth (in microns) within the glue, with the end of the freestanding sample at $x=0$.

	We take as our initial conditions that the entire mesh is at a constant temperature $T_0$.
	The linearity of the heat equation allows us to set this temperature as $T_0=0$ without loss of generality, interpreting the reported temperature as the deviation from this reference.
	We employ Dirichlet boundary conditions enforcing $T=T_0$ on three surfaces; the bottom faces of both of the titanium mounting blocks, and the far end of the thermocouple wires, 2~mm away from the center of the sample.
	All other surfaces are assumed to be thermally insulated.

	\begin{widetext}
	We employ an implicit Crank-Nicholson trapezoidal scheme to advance the heat equation in time.
	We calculate the temperature distribution at the $(n+1)$\th{} timestep by solving
	\begin{equation}\label{eq:timestep}
		c(\vec{x})\frac{T_{n+1}-T_n}{\Delta t} = \frac{\partial}{\partial x_i}\left[k_{ij}(\vec{x})\frac{\partial}{\partial x_j}\left(\frac{T_{n+1}+T_n}{2}\right)\right] + \frac{Q_{n+1}+Q_n}{2}
	\end{equation}
	which is then converted into the weak variational form through standard techniques.\cite{Langtangen2016}
	We use a space of linear basis functions defined on scalar elements.
	\end{widetext}

	In order to extract the magnitude and phase of the thermal transfer function from the temperature profile, we must first run the simulation through a finite number of cycles.
	For all simulations shown here, the time step $\Delta t$ was chosen such that $\Delta t = 1/(100f)$, where $f$ is the strain frequency.
	At each frequency, we allow the system to run through ten full cycles, or 1000 timesteps, extracting the temperature from various relevant locations within the sample and thermometer at each step.
	To minimize the effect of transient fluctuations in the finite number of cycles, we discard the results of the first five cycles, and perform a sinusoidal fit to the final five cycles.

\section{Results of FEM simulations}\label{sec:results}

	\subsection{Sample-bath coupling}\label{sec:bath}
		\begin{figure*}
			\centering
			\includegraphics[width=\textwidth]{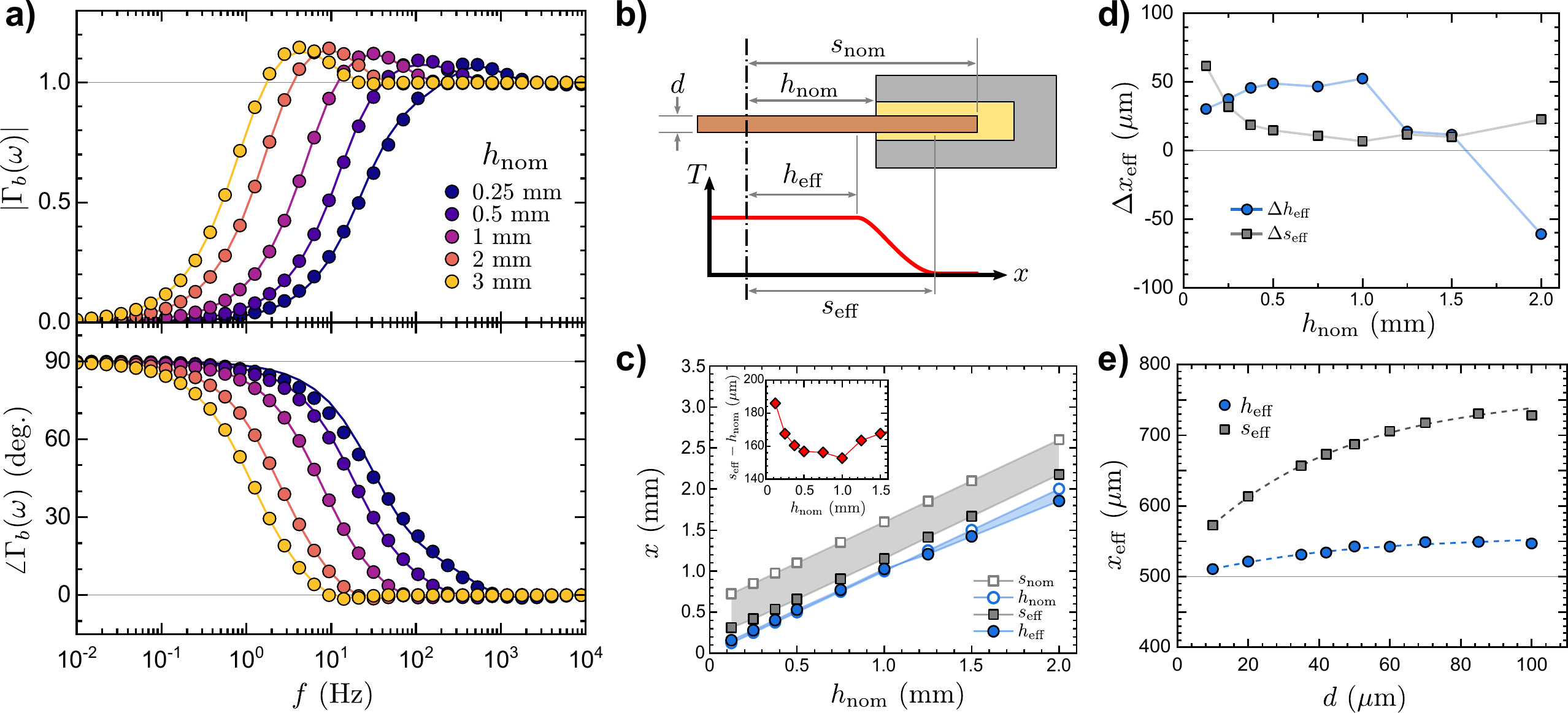}
			\caption[Results of FEM simulations of the effects of mounting plates and sample geometry on the AC-ECE sensitivity function]{Results of FEM simulations of the effects of mounting plates and sample geometry on the AC-ECE sensitivity function.
			(a) Magnitude and phase of ECE temperature oscillations within the sample for different values of plate spacing $h_\mathrm{nom}$.
			Filled symbols correspond to numerical results, and solid lines represent fits of the continuum model of \cref{eq:contgamma} using the lengths $s_\mathrm{eff}$ and $h_\mathrm{eff}$ as free parameters.
			(b) Schematic defining the nominal and effective sample dimensions.
			(c) Dependence of the effective sample dimensions on the sample length.
			Open symbols show the nominal values, and shaded regions highlight the observed deviations.
			Inset to (c) shows the relatively weak dependence of the thermalization length of the sample, defined as $s_\mathrm{eff}-h_\mathrm{nom}$ as a function of sample length.
			(d) Differences in effective dimensions with and without including the inhomogeneous strain within the glued sections of the sample, showing the relatively small role the details of strain relaxation within the sample ends plays in the overall response.
			(e) Effective dimensions as a function of sample thickness.
			Dashed lines indicate exponential fits to the data as described in the text.
			}
			\label{fig:mountplates}
		\end{figure*}

	We begin by studying the details of the thermal connection between the sample and the reservoir.
	To again isolate only the $\Gamma_b(\omega)$ component, we remove the thermocouple and bonding material from the geometry shown in \cref{fig:ecgeo}.
	We then generate several instances of the mesh with varying sample length and thickness.
	For all of the calculations in this section we maintain the same dimensions of the mounting plate blocks, the width of the sample, and the length of the glued regions of the sample.

	The sample itself can be expected to behave in a similar fashion to what was presented in Section \ref{sec:continuum}, although the boundary conditions differ slightly near the ends.
	We define the nominal total sample length (including the glued regions) as $2s_\mathrm{nom}$, and the spacing between the mounting plates as $2h_\mathrm{nom}$.
	To account for the differences in boundary conditions, we also define effective dimensions $s_\mathrm{eff}$ and $h_\mathrm{eff}$, for which \cref{eq:contgamma} reproduces the behavior most faithfully.
	These parameters are defined schematically in \cref{fig:mountplates}(b).
	These two effective dimensions are the only free parameters in a least-squares fit of \cref{eq:contgamma} to the FEM results.

	The results of FEM calculations for several different mounting plate spacings are shown in \cref{fig:mountplates}(a), superimposed over fits to the continuum model.

	Good fits are achieved for all sample lengths, with the shortest samples displaying the largest deviations.
	The frequency of the crossover peak scales with the inverse square of the sample length, as would be expected from the diffusion equation.
	The initial rise of the magnitude also varies in steepness, becoming more gradual as the sample decreases in length.
	This change in shape is also observed in \cref{fig:continuum}(a), as a consequence of changing the $h/s$ ratio.

	The nominal and effective lengths extracted from fits to the FEM results are presented in \cref{fig:mountplates}(c).
	As $h_\mathrm{nom}$ increases, the effective dimensions grow approximately linearly as well.
	The sample thermalization length, defined as the difference between the edge of the mounting plates $h_\mathrm{nom}$ and the total effective half-length of the sample $s_\mathrm{eff}$, is plotted in the inset to \cref{fig:mountplates}(c).
	The thermalization length exhibits only a weak dependence on the plate spacing, adopting values near \SIrange{150}{180}{\micro m} for all spacings studied.

	As mentioned in Section \ref{sec:model}, we have also included the finite strain within the glued regions of the sample in our definition of the heat term $Q$.
	We have performed FEM calculations both with and without these exponential tails, and the difference in the effective dimensions are presented in \cref{fig:mountplates}(d).
	We find relatively very little effect on the resulting effective dimensions, amounting to just a few microns of difference.
	The characteristic strain relaxation length is approximately the same as the sample thermalization length.
	The effective dimensions are most sensitive to strain relaxation effects for small plate spacings, but the total effect is negligible compared to practical uncertainty in the epoxy dimensions and strain transmission.
	As a consequence, the details of strain relaxation within the sample ends are unlikely to have a significant effect on the frequency dependence.

	The operative parameter for controlling the difference between the nominal and effective dimensions is the thickness of the sample.
	As shown in \cref{fig:mountplates}(e), increasing the sample thickness causes an increase in $s_\mathrm{eff}$ and $h_\mathrm{eff}$ following an exponential curve with characteristic length of approximately \SI{42}{\micro m}.
	The growing thermalization length is a consequence of the increasing total thermal conductance, as might be expected for a static thermalization problem\cite{Kopp1971}.

	The effects of the thermal conductivity and heat capacity of the sample mounting epoxy has also been explored.
	We find that varying the diffusivity over three orders of magnitude has very little effect on the resulting effective dimensions.

	The practical consequence of this relatively short thermalization length for the frequency-dependent sensitivity function is that $h_\mathrm{eff}/s_\mathrm{eff}\approx 0.9$ largely independent of the sample dimensions when the sample is of order 1~mm or longer.
	This condition places the response squarely in the regime where the greater-than-unity peak in the transfer function reaches the largest values (cf. inset to \cref{fig:continuum}(b)).

	\subsection{Sample-thermometer coupling}\label{sec:thermo}

		\begin{figure*}
			\includegraphics[width=\textwidth]{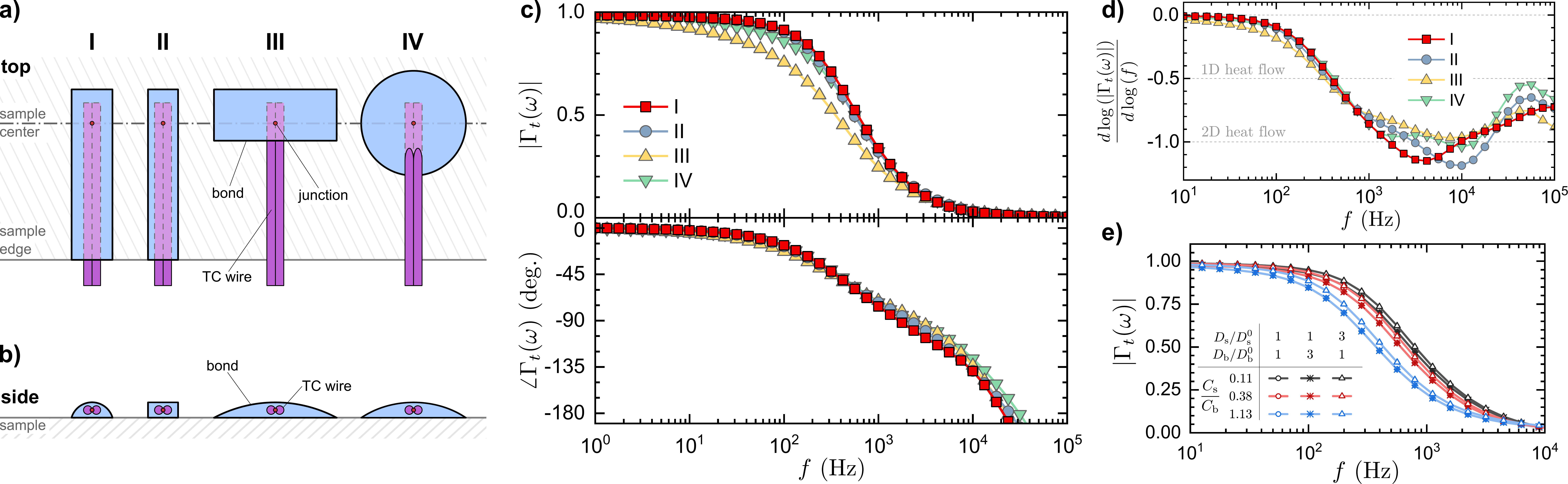}
			\caption[High-frequency contributions to the AC-ECE sensitivity function arising from the geometry and material properties of the thermocouple.]{
				High-frequency contributions to the AC-ECE sensitivity function arising from the geometry and material properties of the thermocouple.
				Four different geometries of the TC bond, each containing the same total volume of the bond material, are shown in panels (a) and (b).
				Panel (c) shows the magnitude and phase response of $\Gamma_t(\omega)$ for each variant.
				We find that bond I (red squares) maintains the largest magnitudes out to the highest frequencies, indicating that a long and thin bond bead with a center of mass as close as possible to the sample surface is the best choice for a given bond volume.
				Panel (d) extracts the effective power law $\alpha = d\log{|\Gamma_t(\omega)}/d\log{(f)}$ describing the frequency dependence of the magnitude.
				For heat flow of dimension $d$, one expects $\alpha=-d_\theta/2$.
				All four thermocouple bond variants initially exhibit a trend in $\alpha$ toward the 2D limit interrupted by a rebound back to $\alpha>-0.5$ over roughly one decade in frequency.
				Panel (e) shows the expected behavior of $|\Gamma_t(\omega)|$ for bond I for various material conditions.
				Increasing the sample diffusivity increases the cutoff frequency most strongly, while the bond diffusivity dominates for $f \gtrsim 10$~kHz.
				Changing the ratio between the sample and bond specific heat (holding the diffusivity constant) causes very little effect.
			}
			\label{fig:thermo}
		\end{figure*}

		We now apply the finite element technique to quantify the high-frequency contribution $\Gamma_t(\omega)$ to the sensitivity function.
		As the sample changes in temperature due to the elastocaloric effect, the thermocouple and bonding material must absorb or release heat in order to change the temperature of the junction.
		This heat flow will suppress the temperature oscillation within the sample, and the magnitude of the suppression is related to the ratio between the heat capacity of the TC $C_\mathrm{TC}$ and the heat capacity of the sample $C_s$.
		Considering the TC as a point particle, the relevant sample volume for calculating $C_s$ is the region of the sample within one thermal diffusion length $L_\mathrm{th}=(D/f)^{1/2}$ of the TC.
		As the frequency $f$ rises and $L_\mathrm{th}$ shrinks below the length, width, and eventually thickness of the sample, the ratio $C_{TC}/C_s$ will vary as $f^{-d_\theta/2}$ where $d_\theta$ is the effective dimensionality of the thermal volume, and $d_\theta$ will increase from 0 to 3.

		In principle, the ideal case would be to directly embed, deposit, or weld the thermocouple to the surface of the sample, minimizing extra thermal mass such as epoxy while creating solid thermal contact.
		A thermocouple which is directly embedded into a sample can react within approximately \SI{10}{\micro s},\cite{Rittel1998} while using the sample surface itself (assuming it is conductive) as one leg of an intrinsic thermocouple can produce response times of \SI{3}{\micro s}\cite{Henning1967}.
		However, use of such techniques will be highly material-dependent and may damage the sample; using a small bead of a bonding material such as an epoxy is a more generally applicable solution.
		Additionally, thermocouples are often adhered to samples by hand.
		This certainly allows for decent measurements to be made, although some variance in the quality of the thermal contact is to be expected.
		In principle, micro-manipulators and small syringes could be employed in order to improve reproducibility.

		In practice, the dimensions of the thermocouple and bonding material may be of similar orders of magnitude to the dimensions of the sample itself.
		In order to test the influence of bond geometry in a controlled way, we have developed a set of four different geometries for the TC bonding material, labeled by Roman numerals I through IV, each of which includes the same volume of material and therefore the same total $C_{TC}$.
		The size, shape and orientation of the thermocouple wires are held fixed, as is the total height of the bonding material above the top surface of the sample.
		The different shapes are shown schematically in \cref{fig:thermo}(a) and (b).
		Bonds~I and II are long and thin volumes running along the thermocouple wire but with cross sections of a circular segment and a rectangle, respectively.
		Bond~III is wide and short, near the TC junction only, and also has a circular segment cross section.
		Bond~IV consists of a spherical cap with its axis of revolution passing through the TC junction.

		We isolate the high frequency contributions $\Gamma_t(\omega)$ of the TC bond by removing the components of the mesh corresponding to the mounting plates and epoxy.
		The entire sample is considered to be strained uniformly at a given frequency and the entire mesh is thermally isolated except for the far end of the TC wires.
		FEM calculations are otherwise identical to those in the previous section.

		Results of FEM calculations for these four meshes are presented in \cref{fig:thermo}(c).
		As expected, the magnitude of $\Gamma_t(\omega)$ approaches unity at low frequencies.
		\footnote{The true value is slightly depressed below one, due to the finite heat capacity of the thermocouple and bond.
		Additionally, the boundary conditions at the far end of the wire will implement $\Gamma_b(\omega)$ effects, but at very low frequencies due to the low total conductance.}

		As the frequency increases, the magnitude drops over approximately one decade of frequency.
		It is observed that for strain frequencies above $\approx 100$~Hz, the sensitivity function is largest for the long, thin, cylindrical bond~I, while the short and wide bond~III performs most poorly.
		The phase of the signal decreases linearly between 100~Hz and 10~kHz, then decreases more quickly at larger frequencies.

		We find that a long thin bead of a bonding material running from the tip of the thermocouple along the wires to the edge of the sample provides better thermal coupling than a shorter, but wider, bead of identical volume and thermal parameters.
		This geometry is a compromise between the optimal case of minimum bond volume and the reality of working with liquid adhesives.
		For example, a low-viscosity epoxy would wick along the surface of the thermocouple wire by capillary action.

		In order to evaluate the effective dimensionality $d_\theta$ of the thermal diffusion volume of the sample, we extract the instantaneous power law dependence of $|\Gamma_t(\omega)|=f^\alpha$ as a logarithmic derivative in \cref{fig:thermo}(d).
		Beginning in the low-frequency limit, $\alpha\approx 0$, indicating that the thermal diffusion volume is larger than the sample.
		As the frequency increases and the thermal length decreases, the magnitude begins to decline rapidly.
		The onset of the decline corresponds to $L_\mathrm{th}\approx$\SI{250}{\micro m}, the length of the bond region.
		Increasing the frequency further, $\alpha$ reaches a minimum just beyond $\alpha=-1$, which corresponds to 2D heat flow.
		This minimum corresponds to the crossover between different thermal bottlenecks: for frequencies below this minimum heat flow is dominated by the changing thermal volume of the sample, while heat flow at higher frequencies is limited by diffusion through the bonding material.
		At this point, the diffusion length within the sample becomes smaller than the short axis of the interfacial area between the sample and the bond.
		For extreme frequencies $\gtrsim 10$~kHz, $\alpha$ indicates an approach to 1D behavior, and the thermal contact between the sample and thermocouple is dominated by flow perpendicular to this interface.

		Finally, we compare the results for different thermal parameters in both the sample and bond material in \cref{fig:thermo}(e).
		We consider changes in the diffusivity of the sample ($D_s$) and the thermocouple bond $D_b$ relative to the default values $D_s^0$ and $D_b^0$, as well as the ratio between the sample and bond heat capacities $C_s/C_b$.
		Of these parameters, we find that the the heat capacity ratio exerts the largest effect on the cutoff frequency $\omega_t$, although an order of magnitude change of $C_s/C_b$ causes $\omega$ to vary by less than a factor of three.
		Changes in the diffusivity of the TC bond material has no noticeable effect, consistent with the interpretation that the behavior is limited by $L_{th}$ within the sample.
		Increasing $D_s$ by a factor of three does increase the cutoff frequency, but only slightly.

		A consequence of the results presented in \cref{fig:thermo}(e) is that \emph{both} the high and low frequency cutoff frequencies (which we will denote by $\omega_t$ and $\omega_b$, respectively) are primarily determined by the geometry and diffusivity of the sample, parameters which cannot generally be controlled by the experimentalist.
		In the context of experiments making use of round-wire thermocouples, then, there are relatively few parameters which can be optimized.
		Further improvements would require a significant paradigm shift, such as the implementation of thin-film thermometers which can be deposited onto the sample surface and the thermal coupling will be very close to ideal for all strain frequencies.
		However, it should be noted that in such a case, strain transmission between the sample and thermometer material will also be nearly perfect, so care must be taken to ensure the thermometer's response is insensitive to strain.

		To widen the quasi-adiabatic region by minimizing $\omega_b$, one can use the longest sample possible.
		Increasing the length by a factor of $\lambda$ will decrease the cutoff frequency by $\lambda^{-2}$.
		Simultaneously, however, the sample strain for a given stress will decrease by $\lambda^{-1}$ and the critical compressive buckling stress will decrease by $\lambda^{-2}$ as well.
		Increasing the sample thickness by a factor of $\gamma$ can counteract the buckling condition somewhat (increasing the buckling force by $\gamma^3$ due to changes in the bending moment of the sample) at the further cost of strain ($\gamma^{-1}$).
		While the practical limits will depend on the material under test, increasing the length, thickness, and driving stress generally provides the best conditions for quasi-adiabatic behavior at the lowest frequencies possible.

		To improve the response on the high-frequency side by maximizing $\omega_t$, the choice of bond material (with the optimum corresponding to minimizing $C_b$, while $D_b$ appears irrelevant) has the largest effect, although even this effect is somewhat muted.
		The shape of the bond bead can alter the cutoff frequency by approximately a factor of two.
		The optimal geometry consists of a thin, low-volume bead which connects a significant portion of the thermocouple wire to the sample surface, and which has a low center of mass relative to the sample surface to maximize thermal coupling.
		Additionally, samples which are thinner than the bond bead is wide, for instance, tend to reduce the cutoff frequency due to a reduction of heat capacity per unit area of the sample.
		However, the cutoff frequency saturates when the sample thickness is increased beyond the width of the bond bead, making sample thickness a poor tuning parameter for increasing the cutoff frequency.

\section{Results of full simulations}\label{sec:data}

	Finally, we combine the low- and high-frequency effects and study the complete sensitivity function $\Gamma(\omega)$ for realistic parameters.

	\subsection{Material property dependence}
	\begin{figure*}
		\centering
		\includegraphics[width=\textwidth]{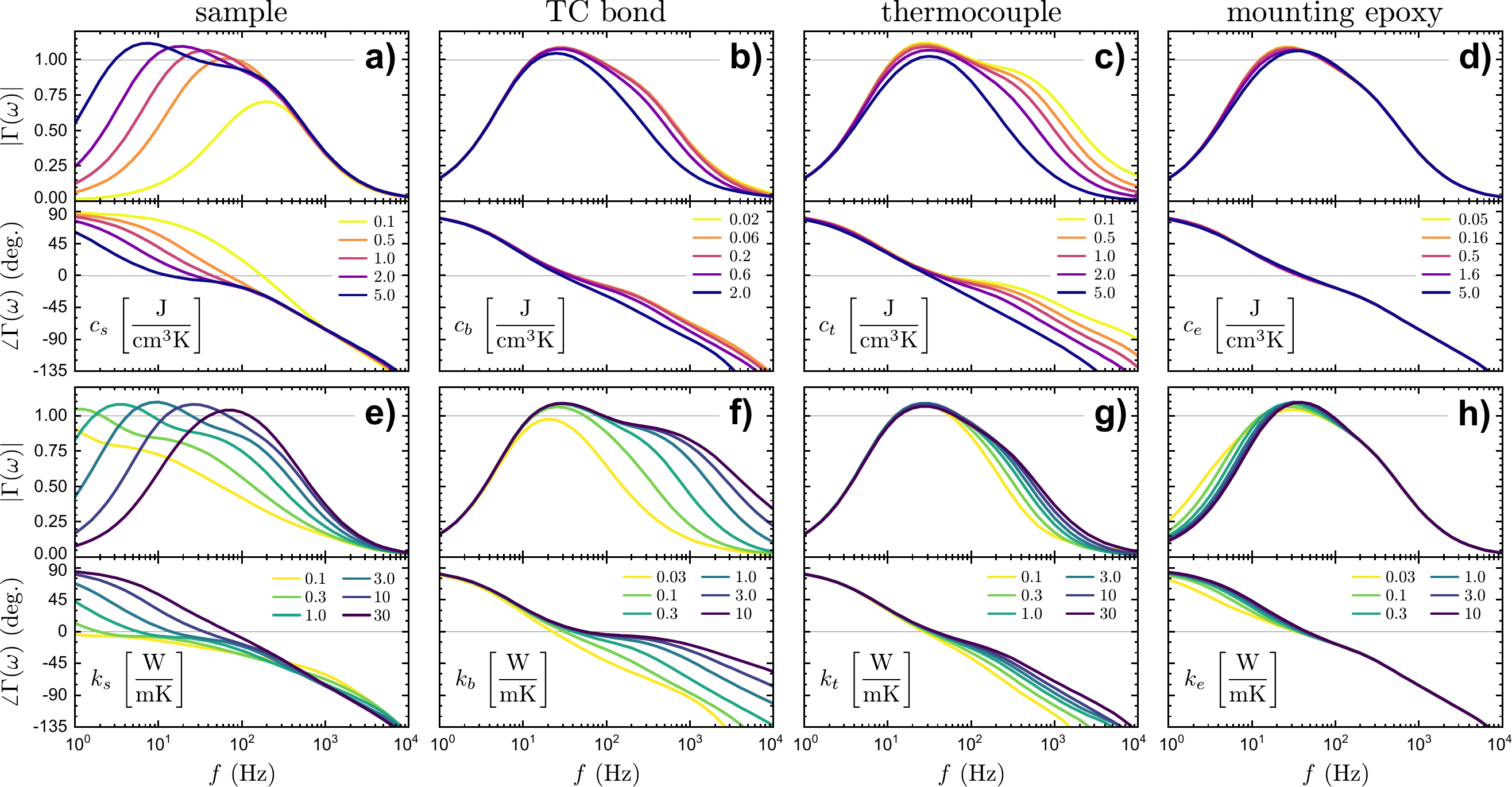}
		\caption[Dependence of the full AC-ECE sensitivity function on frequency and materials parameters]{Dependence of the full AC-ECE sensitivity function $\Gamma(\omega)$ on strain frequency and materials parameters.
			Panels (a)-(d) illustrate the effects of the volumetric heat capacity of the sample, the thermocouple bond, the thermocouple itself, the sample mounting epoxy respectively.
			Panels (e)-(h) show the effect of the thermal conductivity of the same materials.
			Only one parameter is varied at a time; all the rest are held at the values in \cref{tab:params}.
			In panel (e) the sample thermal conductivity anisotropy ratio $k_{xx}/k_{yy}$ is held constant, and the legend corresponds to values of $k_{xx}$.
		}
		\label{fig:materials}
	\end{figure*}

	\Cref{fig:materials} shows a map of $\Gamma(\omega)$ as a function of the heat capacity and thermal conductivity of the sample, sample mounting epoxy, thermocouple, and thermocouple bond material.
	All parameters except the one being varied are set to the values in \cref{tab:params}.
	Despite the variability of the response across the range of parameters, the shape of the response takes on only two qualitatively different forms.
	In the low sensitivity case, $\Gamma(\omega)$ consists of a single sharp peak.
	For larger values of the sensitivity, the response splits into a peak on the low frequency side and a shoulder on the high frequency side.

		\begin{figure*}
			\centering
			\includegraphics[width=\textwidth]{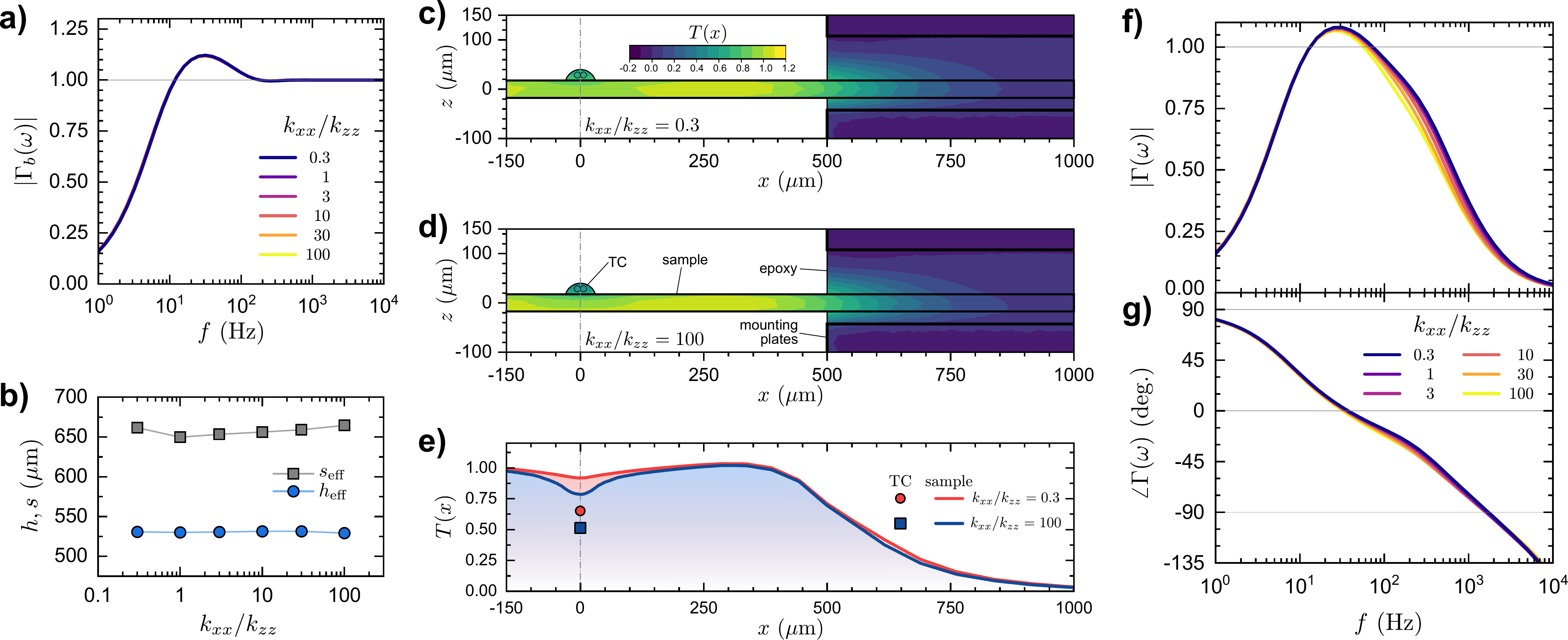}
			\caption[Demonstration of the independence of the AC-ECE sensitivity from the ratio of in-plane to out-of-plane thermal conductivity tensor components $k_{xx}$ and $k_{zz}$.]{Demonstration of the independence of the AC-ECE sensitivity from the ratio of in-plane to out-of-plane thermal conductivity tensor components $k_{xx}$ and $k_{zz}$.
			(a) Magnitude of the bath coupling sensitivity component $\Gamma_b(\omega)$ for a range of anisotropy ratios, holding $k_{xx}$ fixed.
			More than two orders of magnitude of change in $k_{zz}$ causes no noticeable change in $|\Gamma_b(\omega)|$ and all curves collapse onto each other.
			(b) Effective sample dimensions extracted from fits to the continuum model.
			The thermalization lengths are nearly constant.
			(c) Temperature profile for a cross-section of the mesh for $k_{xx}/k_{zz}=0.3$ and (d) $k_{xx}/k_{zz}=0$.
			Both profiles are taken at $f=316.2$~Hz and $t=9.25/f$.
			(e) Temperature extracted running along the top surface of the sample ($z=d/2$) extracted from panels (c) and (d).
			The sample surface temperature is slightly suppressed for the $k_{xx}/k_{zz}=100$ case.
			The instantaneous temperature of the TC junction is shown by the symbols at $x=0$: the measured temperature is suppressed by roughly 0.15 in normalized units for the highly anisotropic case.
			(f) Magnitude and (g) phase of $\Gamma(\omega)$ for FEM simulations of the full assembly.
			While higher anisotropy ratios tend to suppress the high frequency response, the effect is slight.
			}
			\label{fig:aniso}
		\end{figure*}

		Within the field of fundamental condensed matter physics research, many of the most actively studied materials families exhibit significantly anisotropic crystal structures.
		The flat sample morphology presented here is particularly well-suited to layered materials which can be cleaved easily; however, the layered structure usually also implies an anisotropic thermal conductivity as well.
		Thermalization between the sample and thermocouple, as well as between the sample and mounting plates, primarily occurs through heat flow which runs perpendicular to the plane of the sample, whereas thermalization within the sample occurs primarily within the plane.
		As such, anisotropy in the thermal conductivity may have a substantial effect on the sensitivity function $\Gamma(\omega)$.

		We consider a sample material which exhibits a three-, four- or six-fold symmetry axis normal to the plane (parallel to the $\hat{z}$-axis), such that the thermal conductivity tensor $k_{ij}$ takes on the form
		\begin{equation}
			k_ij = \begin{pmatrix}
				k_{xx}	&	0	    	&	0	\\
				0				&	k_{xx}	&	0	\\
				0				&					&	k_{zz}
			\end{pmatrix}.
		\end{equation}
		We neglect any strain-induced anisotropy within the plane.
		A series of FEM calculations varying the out-of-plane component $k_{zz}$ while holding $k_{xx}$ fixed is presented in \cref{fig:aniso}.
		Focusing first on the behavior of the bath coupling component $\Gamma_b(\omega)$, we see that increasing $k_{zz}$ has almost no change in the magnitude.
		Fits of this data to the continuum model of \cref{eq:contgamma} show
		that neither of the effective lengths depend on the out-of-plane thermal conductivity, further indicating that $z$-axis heat flow normal to the the sample-epoxy interface is significantly less important than the $x$-axis heat flow between the strained and unstrained sections of the sample.

		The temperature measured at the thermocouple junction, however, does depend modestly on $k_{zz}$.
		Temperature profiles in the cross-section of the sample shown in \cref{fig:aniso}(c)-(e) show that the actual temperature within the sample hardly varies only slightly.
		The behavior is most different on the top surface of the sample, where it is suppressed slightly in the highly anisotropic case.
		The thermocouple temperature is similarly affected.
		The resulting impact of this effect on the sensitivity function is a slight decrease in the magnitude on the high frequency side of the peak, but the difference is minor.
		In the case of an extremely thermally anisotropic material, If shaping a sample such that the largest thermal conductivity points normal to the plane is not possible, then the frequency range can be improved slightly by adhering the thermocouple to the side of the sample rather than the top.

	\subsection{Comparisons to experiment}
		\begin{figure}
			\centering
			\includegraphics[width=\columnwidth]{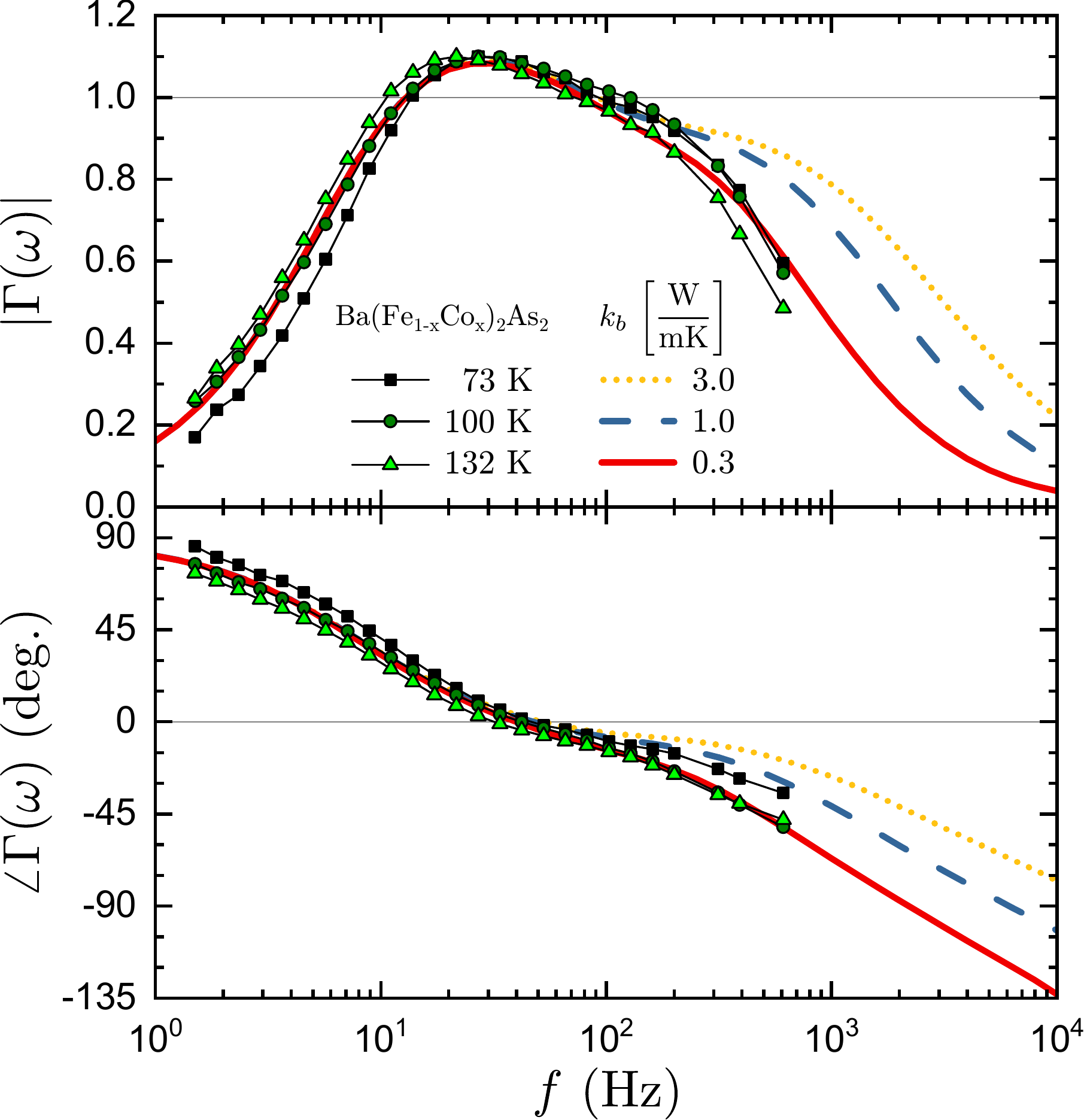}
			\caption[Comparison between experimental results of the AC-ECE technique on \bafeas{} and the results of full FEM simulation]{Comparison between experimental results of the AC-ECE technique on \bafeas{} and the results of full FEM simulation.
			Two free parameters are needed to acquire quantitative agreement--the thermal conductivity of the TC bonding material $k_b$, and the peak magnitude of the elastocaloric effect.
			All other parameters are set to the independently verified values in \cref{tab:params}.
			The experimental data has been scaled using the quantification technique presented in this section, and the quantitative match between the simulation and experiment indicates that this measurement faithfully reproduced the intrinsic elastocaloric tensor $\eta_{ij}$.}
			\label{fig:data}
		\end{figure}

		We now revisit the experimental results for AC-ECE measurements on Co-doped BaFe$_2$As$_2$ and compare the phenomenology to FEM results for the full mesh presented in	 \cref{fig:ecgeo}.
		A comparison between experimental results at three different temperatures are compared to three values of the thermal conductivity of the TC bonding material.
		Unlike the simplified model presented in \cref{fig:intro}, the FEM results accurately predict the shape of the signal.

		Two parameters were required to bring the experimental and simulated curves into quantitative agreement.
		One is the vertical scale factor between the measured elastocaloric effect and $|\Gamma(\omega)|$, which is the quantity we are trying to measure.
		The other parameter is the thermal conductivity of the bonding material.
		Depending on the type of bond used, the thermal properties or the shape of the bond may be unknown or poorly defined.
		Allowing one of the thermal properties of this material to vary, however, is sufficient to capture all of these effects.
		Most importantly, the elastocaloric scale factor and the thermal properties of the bond affect the sensitivity function in orthogonal ways--the scale factor only affects the height of the curve while the thermal parameters affect the width of the response in frequency space.

		\begin{figure}
			\centering
			\includegraphics[width=\columnwidth]{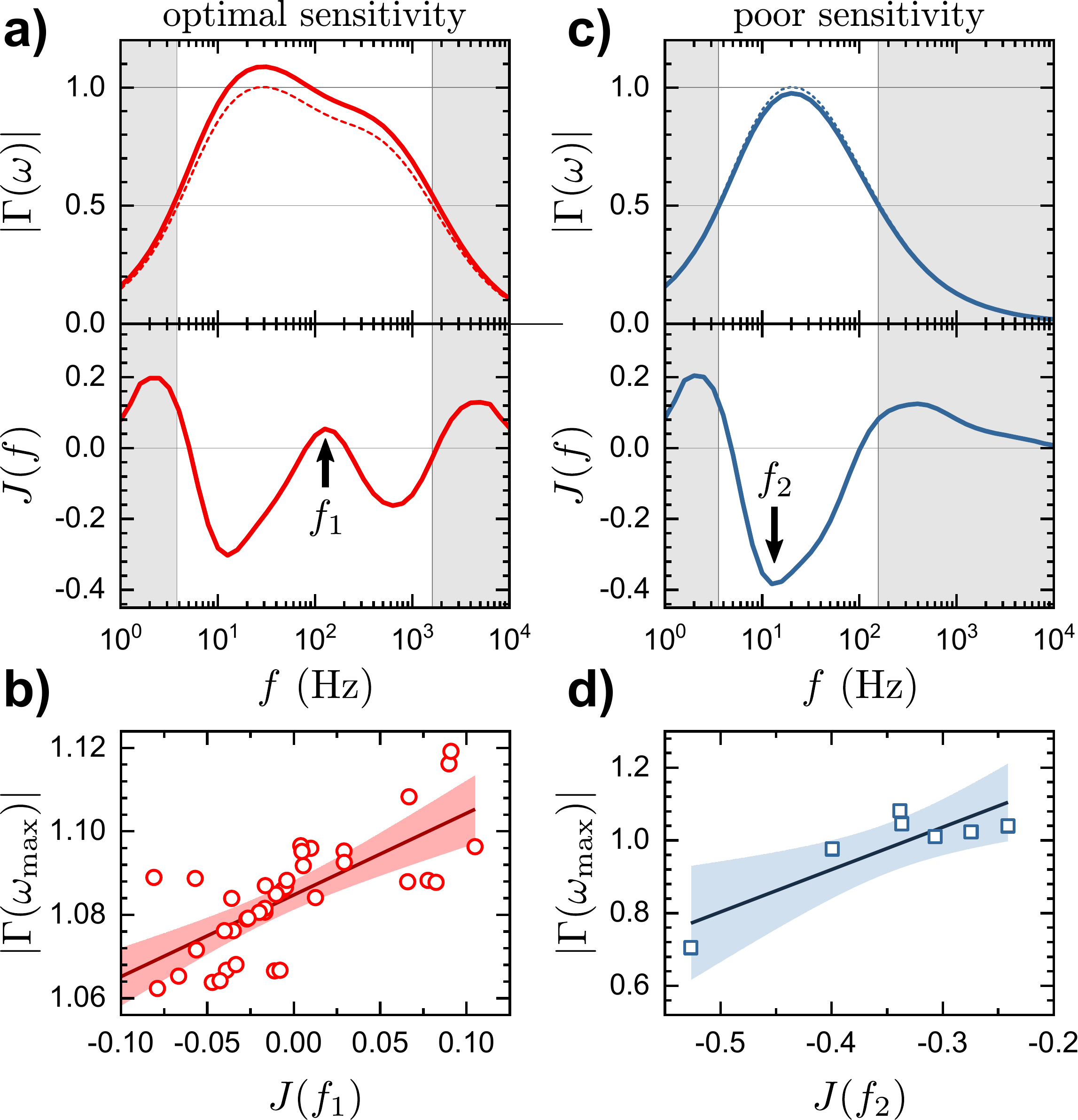}
			\caption[Graphical depiction of an empirical method for estimating the peak sensitivity of the AC-ECE technique based on the shape of the frequency dependence]{Graphical depiction of an empirical method for estimating the peak sensitivity of the AC-ECE technique based on the shape of the frequency dependence.
			Panels (a) and (b) show a representative simulation for which the sensitivity is high, and (c) and (d) show the case of poor sensitivity.
			Considering only data within the full-width at half-maximum of the normalized sensitivity, calculate $J(f)$, the second derivative with respect to $\log{(f)}$ as described in the text.
			If $J(f)$ exhibits a local maximum at $f_1$ as in (a), the approximate peak magnitude can be read from panel (b) or calculated using \cref{eq:quantf1}.
			If no such peak exists, the peak sensitivity can be estimated based on the minimum at $f_2$ using panel (d) or \cref{eq:quantf2}.
			Shaded regions in (b) and (d) indicate 95\% confidence intervals of a linear fit to values extracted from the curves in \cref{fig:materials}.
			}
			\label{fig:quant}
		\end{figure}

		In practice, when $\eta_{ij}$ is not known \textit{a priori}, the frequency dependence of the signal can be used to determine the quality of the thermal coupling for a given sample and material.
		Unlike specific heat measurements in which an external heater is used\cite{Velichkov1992}, a flat plateau is almost never to be expected in an AC-ECE measurement implemented using round wires.
		Instead, we propose an empirical prescription for evaluating the quality of a given measurement and the peak magnitude of $\Gamma(\omega)$ without requiring recourse to one's own finite element calculations.
		The prescription is shown graphically in \cref{fig:quant}.

		Consider a single measurement of the frequency-dependent AC-ECE signal, consisting of oscillations in temperature $T^\prime(\omega)$ and strain $\varepsilon_{ij}^\prime(\omega)$, where both are complex quantities.
		In what follows, we denote measured quantities with primes, while unprimed variables denote the true value corrected for frequency-dependence of the sensitivity.
		Calculate the raw elastocaloric response $\eta_{ij}^\prime(\omega) = T^\prime(\omega)/\varepsilon_{ij}^\prime(\omega)$ calculate the normalized magnitude $H(f)$ defined as
		\begin{equation}
			H(f) = \frac{|\eta_{ij}^\prime(\omega)|}{|\eta_{ij}^\prime(\omega_\mathrm{max}|}.
		\end{equation}
		where $\omega=2\pi f$.
		Define a concavity function $J(\omega)$ which is the second derivative of $H(f)$ with respect to $\log{(f)}$
		\begin{equation}
			J(f) = \frac{d^2 H(f)}{d(\log{(f)})^2}
		\end{equation}

		In ``optimal'' cases, where the experimental setup permits $|\Gamma(\omega_\mathrm{max})|\approx 1.1$, $J(f)$ will exhibit a local maximum at $f_1$.
		By calculating $J(f)$ for all of the simulated traces in \cref{fig:materials} and comparing its behavior to the simulated $\Gamma(\omega)$, we have found consistent relationships between the two despite the variability in terms of frequency dependence.
		By comparing the value of $J$ at this local maximum to the true peak in the sensitivity function $|\Gamma(\omega_\mathrm{max})|$, we have found the two to be related linearly by the equation
		\begin{equation}\label{eq:quantf1}
			|\Gamma(\omega_\mathrm{max})| = 1.085(2) + 0.20(3) \cdot J(f_1).
		\end{equation}
		In cases where no such local maximum is observed, extract the local minimum of $J$ at $f_2$, which lies on the low frequency side of the peak in $H$.
		in this case, the peak sensitivity can be estimated with the relation
		\begin{equation}\label{eq:quantf2}
			|\Gamma(\omega_\mathrm{max})| = 1.39(10) + 1.16(30) J(f_2).
		\end{equation}
		Regardless of which case is used, the peak sensitivity $|\Gamma(\omega_\mathrm{max})|$ sets the scale factor for calculating the bare elastocaloric tensor.
		\begin{equation}
			\eta_{ij} = \frac{\eta_{ij}^\prime(\omega_\mathrm{max}) }{ |\Gamma(\omega_\mathrm{max})|}
		\end{equation}
		This procedure assumes that the bare elastocaloric tensor is frequency independent.

	The results of this study clearly demonstrate both the possibility of and protocol for measuring the AC-ECE in absolute units.
	However, even once the experiment has been optimized it should be remembered that $\Gamma(\omega)$ will vary with temperature as the thermal parameters shift.
	The sloping ``pseudo-plateau'' region of \cref{fig:quant}(a), for example, spans approximately 25\% of the maximum sensitivity.
	Quantitative measurements of $\eta_{ij}$ as a function of temperature therefore require that the frequency dependence be measured for a range of temperatures of interest in order to correct for these effects.

\section{Conclusions}
	Numerical and analytical heat flow studies have been applied to the frequency-dependent AC elastocaloric sensitivity function $\Gamma(\omega)=\Gamma_b(\omega)\Gamma_t(\omega)$.
	We have shown through analytical models that coupling between the sample and the bath always give rise to a small peak where $|\Gamma_b(\omega)|>1$, which arises as a consequence of the finite spatial extent of the elastocalorically excited region.
	By comparison to finite element simulation, we have also quantified the effects of sample dimensions and strain relaxation.
	Examination of the decoupling behavior of the thermocouple at high frequencies indicates that the the optimal thermocouple bond geometry is a long, thin bead with a center of mass closest to the sample surface.
	We have demonstrated the effects of the heat capacity and thermal conductivity of the various materials involved in this measurement, including anisotropy in the sample thermal conductivity tensor.
	By combining both the high and low frequency components we have shown good agreement with data acquired on a sample of \bafeas{}.
	Finally, we have provided an empirical technique for estimating the absolute magnitude of the elastocaloric tensor through measurements of the frequency dependence.
	This work provides an intuitive baseline for the detailed interpretation of the frequency dependence of AC elastocaloric effect measurements.

\section{Acknowledgments}
	The authors wish to thank Phil Walmsley for insightful discussions.
	This work was supported by the U. S. Department of Energy (DOE) Office of Basic Energy Science, Division of Materials Science and Engineering at Stanford under contract No. DE-AC02-76SF00515.

\bibliographystyle{apsrev4-1}
\bibliography{ElastoCaloricFEM}

\begin{thebibliography}{42}%
\makeatletter
\providecommand \@ifxundefined [1]{%
 \@ifx{#1\undefined}
}%
\providecommand \@ifnum [1]{%
 \ifnum #1\expandafter \@firstoftwo
 \else \expandafter \@secondoftwo
 \fi
}%
\providecommand \@ifx [1]{%
 \ifx #1\expandafter \@firstoftwo
 \else \expandafter \@secondoftwo
 \fi
}%
\providecommand \natexlab [1]{#1}%
\providecommand \enquote  [1]{``#1''}%
\providecommand \bibnamefont  [1]{#1}%
\providecommand \bibfnamefont [1]{#1}%
\providecommand \citenamefont [1]{#1}%
\providecommand \href@noop [0]{\@secondoftwo}%
\providecommand \href [0]{\begingroup \@sanitize@url \@href}%
\providecommand \@href[1]{\@@startlink{#1}\@@href}%
\providecommand \@@href[1]{\endgroup#1\@@endlink}%
\providecommand \@sanitize@url [0]{\catcode `\\12\catcode `\$12\catcode
  `\&12\catcode `\#12\catcode `\^12\catcode `\_12\catcode `\%12\relax}%
\providecommand \@@startlink[1]{}%
\providecommand \@@endlink[0]{}%
\providecommand \url  [0]{\begingroup\@sanitize@url \@url }%
\providecommand \@url [1]{\endgroup\@href {#1}{\urlprefix }}%
\providecommand \urlprefix  [0]{URL }%
\providecommand \Eprint [0]{\href }%
\providecommand \doibase [0]{http://dx.doi.org/}%
\providecommand \selectlanguage [0]{\@gobble}%
\providecommand \bibinfo  [0]{\@secondoftwo}%
\providecommand \bibfield  [0]{\@secondoftwo}%
\providecommand \translation [1]{[#1]}%
\providecommand \BibitemOpen [0]{}%
\providecommand \bibitemStop [0]{}%
\providecommand \bibitemNoStop [0]{.\EOS\space}%
\providecommand \EOS [0]{\spacefactor3000\relax}%
\providecommand \BibitemShut  [1]{\csname bibitem#1\endcsname}%
\let\auto@bib@innerbib\@empty
\bibitem [{\citenamefont {Zi{\'{o}}tkowski}(1993)}]{Ziotkowski1993}%
  \BibitemOpen
  \bibfield  {author} {\bibinfo {author} {\bibfnamefont {A.}~\bibnamefont
  {Zi{\'{o}}tkowski}},\ }\href@noop {} {\bibfield  {journal} {\bibinfo
  {journal} {Mech. Mater.}\ }\textbf {\bibinfo {volume} {16}},\ \bibinfo
  {pages} {365} (\bibinfo {year} {1993})}\BibitemShut {NoStop}%
\bibitem [{\citenamefont {Cui}\ \emph {et~al.}(2012)\citenamefont {Cui},
  \citenamefont {Wu}, \citenamefont {Muehlbauer}, \citenamefont {Hwang},
  \citenamefont {Radermacher}, \citenamefont {Fackler}, \citenamefont
  {Wuttig},\ and\ \citenamefont {Takeuchi}}]{Cui2012}%
  \BibitemOpen
  \bibfield  {author} {\bibinfo {author} {\bibfnamefont {J.}~\bibnamefont
  {Cui}}, \bibinfo {author} {\bibfnamefont {Y.}~\bibnamefont {Wu}}, \bibinfo
  {author} {\bibfnamefont {J.}~\bibnamefont {Muehlbauer}}, \bibinfo {author}
  {\bibfnamefont {Y.}~\bibnamefont {Hwang}}, \bibinfo {author} {\bibfnamefont
  {R.}~\bibnamefont {Radermacher}}, \bibinfo {author} {\bibfnamefont
  {S.}~\bibnamefont {Fackler}}, \bibinfo {author} {\bibfnamefont
  {M.}~\bibnamefont {Wuttig}}, \ and\ \bibinfo {author} {\bibfnamefont
  {I.}~\bibnamefont {Takeuchi}},\ }\href@noop {} {\bibfield  {journal}
  {\bibinfo  {journal} {Appl. Phys. Lett.}\ }\textbf {\bibinfo {volume}
  {101}},\ \bibinfo {pages} {2} (\bibinfo {year} {2012})}\BibitemShut {NoStop}%
\bibitem [{\citenamefont {Qian}\ \emph
  {et~al.}(2016{\natexlab{a}})\citenamefont {Qian}, \citenamefont {Geng},
  \citenamefont {Wang}, \citenamefont {Ling}, \citenamefont {Hwang},
  \citenamefont {Radermacher}, \citenamefont {Takeuchi},\ and\ \citenamefont
  {Cui}}]{Qian2016}%
  \BibitemOpen
  \bibfield  {author} {\bibinfo {author} {\bibfnamefont {S.}~\bibnamefont
  {Qian}}, \bibinfo {author} {\bibfnamefont {Y.}~\bibnamefont {Geng}}, \bibinfo
  {author} {\bibfnamefont {Y.}~\bibnamefont {Wang}}, \bibinfo {author}
  {\bibfnamefont {J.}~\bibnamefont {Ling}}, \bibinfo {author} {\bibfnamefont
  {Y.}~\bibnamefont {Hwang}}, \bibinfo {author} {\bibfnamefont
  {R.}~\bibnamefont {Radermacher}}, \bibinfo {author} {\bibfnamefont
  {I.}~\bibnamefont {Takeuchi}}, \ and\ \bibinfo {author} {\bibfnamefont
  {J.}~\bibnamefont {Cui}},\ }\href
  {http://dx.doi.org/10.1016/j.ijrefrig.2015.12.001
  https://linkinghub.elsevier.com/retrieve/pii/S0140700715003783} {\bibfield
  {journal} {\bibinfo  {journal} {Int. J. Refrig.}\ }\textbf {\bibinfo {volume}
  {64}},\ \bibinfo {pages} {1} (\bibinfo {year}
  {2016}{\natexlab{a}})}\BibitemShut {NoStop}%
\bibitem [{\citenamefont {Qian}\ \emph
  {et~al.}(2016{\natexlab{b}})\citenamefont {Qian}, \citenamefont {Nasuta},
  \citenamefont {Rhoads}, \citenamefont {Wang}, \citenamefont {Geng},
  \citenamefont {Hwang}, \citenamefont {Radermacher},\ and\ \citenamefont
  {Takeuchi}}]{Qian2016a}%
  \BibitemOpen
  \bibfield  {author} {\bibinfo {author} {\bibfnamefont {S.}~\bibnamefont
  {Qian}}, \bibinfo {author} {\bibfnamefont {D.}~\bibnamefont {Nasuta}},
  \bibinfo {author} {\bibfnamefont {A.}~\bibnamefont {Rhoads}}, \bibinfo
  {author} {\bibfnamefont {Y.}~\bibnamefont {Wang}}, \bibinfo {author}
  {\bibfnamefont {Y.}~\bibnamefont {Geng}}, \bibinfo {author} {\bibfnamefont
  {Y.}~\bibnamefont {Hwang}}, \bibinfo {author} {\bibfnamefont
  {R.}~\bibnamefont {Radermacher}}, \ and\ \bibinfo {author} {\bibfnamefont
  {I.}~\bibnamefont {Takeuchi}},\ }\href
  {http://dx.doi.org/10.1016/j.ijrefrig.2015.10.019
  https://linkinghub.elsevier.com/retrieve/pii/S014070071500314X} {\bibfield
  {journal} {\bibinfo  {journal} {Int. J. Refrig.}\ }\textbf {\bibinfo {volume}
  {62}},\ \bibinfo {pages} {177} (\bibinfo {year}
  {2016}{\natexlab{b}})}\BibitemShut {NoStop}%
\bibitem [{\citenamefont {Luo}\ \emph {et~al.}(2017)\citenamefont {Luo},
  \citenamefont {Feng},\ and\ \citenamefont {Verma}}]{Luo2017}%
  \BibitemOpen
  \bibfield  {author} {\bibinfo {author} {\bibfnamefont {D.}~\bibnamefont
  {Luo}}, \bibinfo {author} {\bibfnamefont {Y.}~\bibnamefont {Feng}}, \ and\
  \bibinfo {author} {\bibfnamefont {P.}~\bibnamefont {Verma}},\ }\href
  {http://dx.doi.org/10.1016/j.energy.2017.05.008
  https://linkinghub.elsevier.com/retrieve/pii/S0360544217307466} {\bibfield
  {journal} {\bibinfo  {journal} {Energy}\ }\textbf {\bibinfo {volume} {130}},\
  \bibinfo {pages} {500} (\bibinfo {year} {2017})}\BibitemShut {NoStop}%
\bibitem [{\citenamefont {Chu}\ \emph {et~al.}(2012)\citenamefont {Chu},
  \citenamefont {Kuo}, \citenamefont {Analytis},\ and\ \citenamefont
  {Fisher}}]{Chu2012}%
  \BibitemOpen
  \bibfield  {author} {\bibinfo {author} {\bibfnamefont {J.}~\bibnamefont
  {Chu}}, \bibinfo {author} {\bibfnamefont {H.}~\bibnamefont {Kuo}}, \bibinfo
  {author} {\bibfnamefont {J.~G.}\ \bibnamefont {Analytis}}, \ and\ \bibinfo
  {author} {\bibfnamefont {I.~R.}\ \bibnamefont {Fisher}},\ }\href
  {http://www.sciencemag.org/content/337/6095/710.abstract
  http://www.sciencemag.org/cgi/doi/10.1126/science.1221713} {\bibfield
  {journal} {\bibinfo  {journal} {Science}\ }\textbf {\bibinfo {volume}
  {337}},\ \bibinfo {pages} {710} (\bibinfo {year} {2012})}\BibitemShut
  {NoStop}%
\bibitem [{\citenamefont {Hicks}\ \emph
  {et~al.}(2014{\natexlab{a}})\citenamefont {Hicks}, \citenamefont {Brodsky},
  \citenamefont {Yelland}, \citenamefont {Gibbs}, \citenamefont {Bruin},
  \citenamefont {Barber}, \citenamefont {Edkins}, \citenamefont {Nishimura},
  \citenamefont {Yonezawa}, \citenamefont {Maeno},\ and\ \citenamefont
  {Mackenzie}}]{Hicks2014a}%
  \BibitemOpen
  \bibfield  {author} {\bibinfo {author} {\bibfnamefont {C.~W.}\ \bibnamefont
  {Hicks}}, \bibinfo {author} {\bibfnamefont {D.~O.}\ \bibnamefont {Brodsky}},
  \bibinfo {author} {\bibfnamefont {E.~A.}\ \bibnamefont {Yelland}}, \bibinfo
  {author} {\bibfnamefont {A.~S.}\ \bibnamefont {Gibbs}}, \bibinfo {author}
  {\bibfnamefont {J.~A.~N.}\ \bibnamefont {Bruin}}, \bibinfo {author}
  {\bibfnamefont {M.~E.}\ \bibnamefont {Barber}}, \bibinfo {author}
  {\bibfnamefont {S.~D.}\ \bibnamefont {Edkins}}, \bibinfo {author}
  {\bibfnamefont {K.}~\bibnamefont {Nishimura}}, \bibinfo {author}
  {\bibfnamefont {S.}~\bibnamefont {Yonezawa}}, \bibinfo {author}
  {\bibfnamefont {Y.}~\bibnamefont {Maeno}}, \ and\ \bibinfo {author}
  {\bibfnamefont {A.~P.}\ \bibnamefont {Mackenzie}},\ }\href
  {https://www.sciencemag.org/lookup/doi/10.1126/science.1248292} {\bibfield
  {journal} {\bibinfo  {journal} {Science}\ }\textbf {\bibinfo {volume}
  {344}},\ \bibinfo {pages} {283} (\bibinfo {year}
  {2014}{\natexlab{a}})}\BibitemShut {NoStop}%
\bibitem [{\citenamefont {Kim}\ \emph {et~al.}(2018)\citenamefont {Kim},
  \citenamefont {Souliou}, \citenamefont {Barber}, \citenamefont
  {Lefran{\c{c}}ois}, \citenamefont {Minola}, \citenamefont {Tortora},
  \citenamefont {Heid}, \citenamefont {Nandi}, \citenamefont {Borzi},
  \citenamefont {Garbarino}, \citenamefont {Bosak}, \citenamefont {Porras},
  \citenamefont {Loew}, \citenamefont {K{\"{o}}nig}, \citenamefont {Moll},
  \citenamefont {Mackenzie}, \citenamefont {Keimer}, \citenamefont {Hicks},\
  and\ \citenamefont {{Le Tacon}}}]{Kim2018a}%
  \BibitemOpen
  \bibfield  {author} {\bibinfo {author} {\bibfnamefont {H.}~\bibnamefont
  {Kim}}, \bibinfo {author} {\bibfnamefont {S.~M.}\ \bibnamefont {Souliou}},
  \bibinfo {author} {\bibfnamefont {M.~E.}\ \bibnamefont {Barber}}, \bibinfo
  {author} {\bibfnamefont {E.}~\bibnamefont {Lefran{\c{c}}ois}}, \bibinfo
  {author} {\bibfnamefont {M.}~\bibnamefont {Minola}}, \bibinfo {author}
  {\bibfnamefont {M.}~\bibnamefont {Tortora}}, \bibinfo {author} {\bibfnamefont
  {R.}~\bibnamefont {Heid}}, \bibinfo {author} {\bibfnamefont {N.}~\bibnamefont
  {Nandi}}, \bibinfo {author} {\bibfnamefont {R.~A.}\ \bibnamefont {Borzi}},
  \bibinfo {author} {\bibfnamefont {G.}~\bibnamefont {Garbarino}}, \bibinfo
  {author} {\bibfnamefont {A.}~\bibnamefont {Bosak}}, \bibinfo {author}
  {\bibfnamefont {J.}~\bibnamefont {Porras}}, \bibinfo {author} {\bibfnamefont
  {T.}~\bibnamefont {Loew}}, \bibinfo {author} {\bibfnamefont {M.}~\bibnamefont
  {K{\"{o}}nig}}, \bibinfo {author} {\bibfnamefont {P.~J.~W.}\ \bibnamefont
  {Moll}}, \bibinfo {author} {\bibfnamefont {A.~P.}\ \bibnamefont {Mackenzie}},
  \bibinfo {author} {\bibfnamefont {B.}~\bibnamefont {Keimer}}, \bibinfo
  {author} {\bibfnamefont {C.~W.}\ \bibnamefont {Hicks}}, \ and\ \bibinfo
  {author} {\bibfnamefont {M.}~\bibnamefont {{Le Tacon}}},\ }\href
  {https://www.sciencemag.org/lookup/doi/10.1126/science.aat4708} {\bibfield
  {journal} {\bibinfo  {journal} {Science}\ }\textbf {\bibinfo {volume}
  {362}},\ \bibinfo {pages} {1040} (\bibinfo {year} {2018})}\BibitemShut
  {NoStop}%
\bibitem [{\citenamefont {Bachmann}\ \emph {et~al.}(2019)\citenamefont
  {Bachmann}, \citenamefont {Ferguson}, \citenamefont {Theuss}, \citenamefont
  {Meng}, \citenamefont {Putzke}, \citenamefont {Helm}, \citenamefont {Shirer},
  \citenamefont {Li}, \citenamefont {Modic}, \citenamefont {Nicklas},
  \citenamefont {K{\"{o}}nig}, \citenamefont {Low}, \citenamefont {Ghosh},
  \citenamefont {Mackenzie}, \citenamefont {Arnold}, \citenamefont {Hassinger},
  \citenamefont {McDonald}, \citenamefont {Winter}, \citenamefont {Bauer},
  \citenamefont {Ronning}, \citenamefont {Ramshaw}, \citenamefont {Nowack},\
  and\ \citenamefont {Moll}}]{Bachmann2019}%
  \BibitemOpen
  \bibfield  {author} {\bibinfo {author} {\bibfnamefont {M.~D.}\ \bibnamefont
  {Bachmann}}, \bibinfo {author} {\bibfnamefont {G.~M.}\ \bibnamefont
  {Ferguson}}, \bibinfo {author} {\bibfnamefont {F.}~\bibnamefont {Theuss}},
  \bibinfo {author} {\bibfnamefont {T.}~\bibnamefont {Meng}}, \bibinfo {author}
  {\bibfnamefont {C.}~\bibnamefont {Putzke}}, \bibinfo {author} {\bibfnamefont
  {T.}~\bibnamefont {Helm}}, \bibinfo {author} {\bibfnamefont {K.~R.}\
  \bibnamefont {Shirer}}, \bibinfo {author} {\bibfnamefont {Y.~S.}\
  \bibnamefont {Li}}, \bibinfo {author} {\bibfnamefont {K.~A.}\ \bibnamefont
  {Modic}}, \bibinfo {author} {\bibfnamefont {M.}~\bibnamefont {Nicklas}},
  \bibinfo {author} {\bibfnamefont {M.}~\bibnamefont {K{\"{o}}nig}}, \bibinfo
  {author} {\bibfnamefont {D.}~\bibnamefont {Low}}, \bibinfo {author}
  {\bibfnamefont {S.}~\bibnamefont {Ghosh}}, \bibinfo {author} {\bibfnamefont
  {A.~P.}\ \bibnamefont {Mackenzie}}, \bibinfo {author} {\bibfnamefont
  {F.}~\bibnamefont {Arnold}}, \bibinfo {author} {\bibfnamefont
  {E.}~\bibnamefont {Hassinger}}, \bibinfo {author} {\bibfnamefont {R.~D.}\
  \bibnamefont {McDonald}}, \bibinfo {author} {\bibfnamefont {L.~E.}\
  \bibnamefont {Winter}}, \bibinfo {author} {\bibfnamefont {E.~D.}\
  \bibnamefont {Bauer}}, \bibinfo {author} {\bibfnamefont {F.}~\bibnamefont
  {Ronning}}, \bibinfo {author} {\bibfnamefont {B.~J.}\ \bibnamefont
  {Ramshaw}}, \bibinfo {author} {\bibfnamefont {K.~C.}\ \bibnamefont {Nowack}},
  \ and\ \bibinfo {author} {\bibfnamefont {P.~J.}\ \bibnamefont {Moll}},\
  }\href@noop {} {\bibfield  {journal} {\bibinfo  {journal} {Science}\ }\textbf
  {\bibinfo {volume} {366}},\ \bibinfo {pages} {221} (\bibinfo {year}
  {2019})}\BibitemShut {NoStop}%
\bibitem [{\citenamefont {Rosenberg}\ \emph {et~al.}(2019)\citenamefont
  {Rosenberg}, \citenamefont {Chu}, \citenamefont {Ruff}, \citenamefont
  {Hristov},\ and\ \citenamefont {Fisher}}]{Rosenberg2019}%
  \BibitemOpen
  \bibfield  {author} {\bibinfo {author} {\bibfnamefont {E.~W.}\ \bibnamefont
  {Rosenberg}}, \bibinfo {author} {\bibfnamefont {J.}~\bibnamefont {Chu}},
  \bibinfo {author} {\bibfnamefont {J.~P.~C.}\ \bibnamefont {Ruff}}, \bibinfo
  {author} {\bibfnamefont {A.~T.}\ \bibnamefont {Hristov}}, \ and\ \bibinfo
  {author} {\bibfnamefont {I.~R.}\ \bibnamefont {Fisher}},\ }\href
  {http://www.pnas.org/lookup/doi/10.1073/pnas.1818910116} {\bibfield
  {journal} {\bibinfo  {journal} {Proc. Natl. Acad. Sci. U. S. A.}\ }\textbf
  {\bibinfo {volume} {116}},\ \bibinfo {pages} {7232} (\bibinfo {year}
  {2019})}\BibitemShut {NoStop}%
\bibitem [{\citenamefont {Ikeda}\ \emph {et~al.}(2019)\citenamefont {Ikeda},
  \citenamefont {Straquadine}, \citenamefont {Hristov}, \citenamefont
  {Worasaran}, \citenamefont {Palmstrom}, \citenamefont {Sorensen},
  \citenamefont {Walmsley},\ and\ \citenamefont {Fisher}}]{Ikeda2019}%
  \BibitemOpen
  \bibfield  {author} {\bibinfo {author} {\bibfnamefont {M.~S.}\ \bibnamefont
  {Ikeda}}, \bibinfo {author} {\bibfnamefont {J.~A.~W.}\ \bibnamefont
  {Straquadine}}, \bibinfo {author} {\bibfnamefont {A.~T.}\ \bibnamefont
  {Hristov}}, \bibinfo {author} {\bibfnamefont {T.}~\bibnamefont {Worasaran}},
  \bibinfo {author} {\bibfnamefont {J.~C.}\ \bibnamefont {Palmstrom}}, \bibinfo
  {author} {\bibfnamefont {M.}~\bibnamefont {Sorensen}}, \bibinfo {author}
  {\bibfnamefont {P.}~\bibnamefont {Walmsley}}, \ and\ \bibinfo {author}
  {\bibfnamefont {I.~R.}\ \bibnamefont {Fisher}},\ }\href
  {http://aip.scitation.org/doi/10.1063/1.5099924} {\bibfield  {journal}
  {\bibinfo  {journal} {Rev. Sci. Instrum.}\ }\textbf {\bibinfo {volume}
  {90}},\ \bibinfo {pages} {083902} (\bibinfo {year} {2019})}\BibitemShut
  {NoStop}%
\bibitem [{\citenamefont {Hicks}\ \emph
  {et~al.}(2014{\natexlab{b}})\citenamefont {Hicks}, \citenamefont {Barber},
  \citenamefont {Edkins}, \citenamefont {Brodsky},\ and\ \citenamefont
  {Mackenzie}}]{Hicks2014}%
  \BibitemOpen
  \bibfield  {author} {\bibinfo {author} {\bibfnamefont {C.~W.}\ \bibnamefont
  {Hicks}}, \bibinfo {author} {\bibfnamefont {M.~E.}\ \bibnamefont {Barber}},
  \bibinfo {author} {\bibfnamefont {S.~D.}\ \bibnamefont {Edkins}}, \bibinfo
  {author} {\bibfnamefont {D.~O.}\ \bibnamefont {Brodsky}}, \ and\ \bibinfo
  {author} {\bibfnamefont {A.~P.}\ \bibnamefont {Mackenzie}},\ }\href
  {http://dx.doi.org/10.1063/1.4881611} {\bibfield  {journal} {\bibinfo
  {journal} {Rev. Sci. Instrum.}\ }\textbf {\bibinfo {volume} {85}} (\bibinfo
  {year} {2014}{\natexlab{b}})}\BibitemShut {NoStop}%
\bibitem [{\citenamefont {Hristov}\ \emph {et~al.}(2019)\citenamefont
  {Hristov}, \citenamefont {Ikeda}, \citenamefont {Palmstrom},\ and\
  \citenamefont {Fisher}}]{Hristov2019a}%
  \BibitemOpen
  \bibfield  {author} {\bibinfo {author} {\bibfnamefont {A.~T.}\ \bibnamefont
  {Hristov}}, \bibinfo {author} {\bibfnamefont {M.~S.}\ \bibnamefont {Ikeda}},
  \bibinfo {author} {\bibfnamefont {J.~C.}\ \bibnamefont {Palmstrom}}, \ and\
  \bibinfo {author} {\bibfnamefont {I.~R.}\ \bibnamefont {Fisher}},\ }\href
  {http://arxiv.org/abs/1903.04732} {\bibfield  {journal} {\bibinfo  {journal}
  {arXiv}\ } (\bibinfo {year} {2019})},\ \Eprint
  {http://arxiv.org/abs/1903.04732} {arXiv:1903.04732 [cond-mat]} \BibitemShut
  {NoStop}%
\bibitem [{\citenamefont {Ikeda}\ \emph {et~al.}(2018)\citenamefont {Ikeda},
  \citenamefont {Worasaran}, \citenamefont {Palmstrom}, \citenamefont
  {Straquadine}, \citenamefont {Walmsley},\ and\ \citenamefont
  {Fisher}}]{Ikeda2018}%
  \BibitemOpen
  \bibfield  {author} {\bibinfo {author} {\bibfnamefont {M.~S.}\ \bibnamefont
  {Ikeda}}, \bibinfo {author} {\bibfnamefont {T.}~\bibnamefont {Worasaran}},
  \bibinfo {author} {\bibfnamefont {J.~C.}\ \bibnamefont {Palmstrom}}, \bibinfo
  {author} {\bibfnamefont {J.~A.~W.}\ \bibnamefont {Straquadine}}, \bibinfo
  {author} {\bibfnamefont {P.}~\bibnamefont {Walmsley}}, \ and\ \bibinfo
  {author} {\bibfnamefont {I.~R.}\ \bibnamefont {Fisher}},\ }\href
  {https://link.aps.org/doi/10.1103/PhysRevB.98.245133} {\bibfield  {journal}
  {\bibinfo  {journal} {Phys. Rev. B}\ }\textbf {\bibinfo {volume} {98}},\
  \bibinfo {pages} {245133} (\bibinfo {year} {2018})}\BibitemShut {NoStop}%
\bibitem [{\citenamefont {Riou}\ \emph {et~al.}(2004)\citenamefont {Riou},
  \citenamefont {Durastanti},\ and\ \citenamefont {Sfaxi}}]{Riou2004}%
  \BibitemOpen
  \bibfield  {author} {\bibinfo {author} {\bibfnamefont {O.}~\bibnamefont
  {Riou}}, \bibinfo {author} {\bibfnamefont {J.}~\bibnamefont {Durastanti}}, \
  and\ \bibinfo {author} {\bibfnamefont {Y.}~\bibnamefont {Sfaxi}},\ }\href
  {https://linkinghub.elsevier.com/retrieve/pii/S0749603604001181} {\bibfield
  {journal} {\bibinfo  {journal} {Superlattices Microstruct.}\ }\textbf
  {\bibinfo {volume} {35}},\ \bibinfo {pages} {353} (\bibinfo {year}
  {2004})}\BibitemShut {NoStop}%
\bibitem [{Note1()}]{Note1}%
  \BibitemOpen
  \bibinfo {note} {Piezoelectric actuators can, and often are, operated at
  strain frequencies well into the ultrasound range. However, incorporating
  piezoelectric actuators into a uniaxial stress cell assembly\cite {Hicks2014}
  lowers resonance frequencies and creates a risk of vibrational fatigue within
  epoxies and the piezoelectric actuators themselves. Careful design of future
  devices may raise this practical upper bound.}\BibitemShut {Stop}%
\bibitem [{\citenamefont {Hetnarski}(2009)}]{Hetnarski2009}%
  \BibitemOpen
  \bibfield  {author} {\bibinfo {author} {\bibfnamefont {R.~B.}\ \bibnamefont
  {Hetnarski}},\ }\href {http://link.springer.com/10.1007/978-1-4020-9247-3}
  {\emph {\bibinfo {title} {Thermal Stresses – Advanced Theory and
  Applications}}},\ \bibinfo {series} {Solid Mechanics and its Applications},
  Vol.\ \bibinfo {volume} {158}\ (\bibinfo  {publisher} {Springer
  Netherlands},\ \bibinfo {address} {Dordrecht},\ \bibinfo {year}
  {2009})\BibitemShut {NoStop}%
\bibitem [{\citenamefont {Biot}(1956)}]{Biot1956}%
  \BibitemOpen
  \bibfield  {author} {\bibinfo {author} {\bibfnamefont {M.~A.}\ \bibnamefont
  {Biot}},\ }\href {http://aip.scitation.org/doi/10.1063/1.1722351} {\bibfield
  {journal} {\bibinfo  {journal} {J. Appl. Phys.}\ }\textbf {\bibinfo {volume}
  {27}},\ \bibinfo {pages} {240} (\bibinfo {year} {1956})}\BibitemShut
  {NoStop}%
\bibitem [{\citenamefont {Lord}\ and\ \citenamefont
  {Shulman}(1967)}]{Lord1967}%
  \BibitemOpen
  \bibfield  {author} {\bibinfo {author} {\bibfnamefont {H.}~\bibnamefont
  {Lord}}\ and\ \bibinfo {author} {\bibfnamefont {Y.}~\bibnamefont {Shulman}},\
  }\href {https://linkinghub.elsevier.com/retrieve/pii/0022509667900245}
  {\bibfield  {journal} {\bibinfo  {journal} {J. Mech. Phys. Solids}\ }\textbf
  {\bibinfo {volume} {15}},\ \bibinfo {pages} {299} (\bibinfo {year}
  {1967})}\BibitemShut {NoStop}%
\bibitem [{\citenamefont {Green}\ and\ \citenamefont
  {Lindsay}(1972)}]{Green1972}%
  \BibitemOpen
  \bibfield  {author} {\bibinfo {author} {\bibfnamefont {A.~E.}\ \bibnamefont
  {Green}}\ and\ \bibinfo {author} {\bibfnamefont {K.~A.}\ \bibnamefont
  {Lindsay}},\ }\href {http://link.springer.com/10.1007/BF00045689} {\bibfield
  {journal} {\bibinfo  {journal} {J. Elast.}\ }\textbf {\bibinfo {volume}
  {2}},\ \bibinfo {pages} {1} (\bibinfo {year} {1972})}\BibitemShut {NoStop}%
\bibitem [{Note2()}]{Note2}%
  \BibitemOpen
  \bibinfo {note} {Changes of strain imply that work is being done on the
  sample, changing the internal elastic energy even in the adiabatic limit.
  This would also result in a change in the sample temperature; but since both
  compressive and tensile deformations of a solid at equilibrium requires
  positive work, an oscillation at frequency $f$ about the strain-neutral
  condition generates temperature oscillations at frequency $2f$, which will
  not affect phase sensitive measurements. If the oscillation is superimposed
  on a constant strain offset, there \protect \emph {will} be a contribution to
  the signal at $f$ arising from the work done on the sample. Both the
  conventional elastocaloric effect and the elastic energy away from the
  strain-neutral point, however, can be lumped together into our definition of
  $Q(\protect \mathbf {r},t)$.}\BibitemShut {Stop}%
\bibitem [{\citenamefont {Hristov}\ \emph {et~al.}(2018)\citenamefont
  {Hristov}, \citenamefont {Palmstrom}, \citenamefont {Straquadine},
  \citenamefont {Merz}, \citenamefont {Hwang},\ and\ \citenamefont
  {Fisher}}]{Hristov2018}%
  \BibitemOpen
  \bibfield  {author} {\bibinfo {author} {\bibfnamefont {A.~T.}\ \bibnamefont
  {Hristov}}, \bibinfo {author} {\bibfnamefont {J.~C.}\ \bibnamefont
  {Palmstrom}}, \bibinfo {author} {\bibfnamefont {J.~A.~W.}\ \bibnamefont
  {Straquadine}}, \bibinfo {author} {\bibfnamefont {T.~A.}\ \bibnamefont
  {Merz}}, \bibinfo {author} {\bibfnamefont {H.~Y.}\ \bibnamefont {Hwang}}, \
  and\ \bibinfo {author} {\bibfnamefont {I.~R.}\ \bibnamefont {Fisher}},\
  }\href {http://aip.scitation.org/doi/10.1063/1.5031136} {\bibfield  {journal}
  {\bibinfo  {journal} {Rev. Sci. Instrum.}\ }\textbf {\bibinfo {volume}
  {89}},\ \bibinfo {pages} {103901} (\bibinfo {year} {2018})}\BibitemShut
  {NoStop}%
\bibitem [{\citenamefont {Straquadine}\ \emph {et~al.}(2019)\citenamefont
  {Straquadine}, \citenamefont {Palmstrom}, \citenamefont {Walmsley},
  \citenamefont {Hristov}, \citenamefont {Weickert}, \citenamefont {Balakirev},
  \citenamefont {Jaime}, \citenamefont {McDonald},\ and\ \citenamefont
  {Fisher}}]{Straquadine2019a}%
  \BibitemOpen
  \bibfield  {author} {\bibinfo {author} {\bibfnamefont {J.~A.~W.}\
  \bibnamefont {Straquadine}}, \bibinfo {author} {\bibfnamefont {J.~C.}\
  \bibnamefont {Palmstrom}}, \bibinfo {author} {\bibfnamefont {P.}~\bibnamefont
  {Walmsley}}, \bibinfo {author} {\bibfnamefont {A.~T.}\ \bibnamefont
  {Hristov}}, \bibinfo {author} {\bibfnamefont {F.}~\bibnamefont {Weickert}},
  \bibinfo {author} {\bibfnamefont {F.~F.}\ \bibnamefont {Balakirev}}, \bibinfo
  {author} {\bibfnamefont {M.}~\bibnamefont {Jaime}}, \bibinfo {author}
  {\bibfnamefont {R.}~\bibnamefont {McDonald}}, \ and\ \bibinfo {author}
  {\bibfnamefont {I.~R.}\ \bibnamefont {Fisher}},\ }\href
  {https://link.aps.org/doi/10.1103/PhysRevB.100.125147} {\bibfield  {journal}
  {\bibinfo  {journal} {Phys. Rev. B}\ }\textbf {\bibinfo {volume} {100}},\
  \bibinfo {pages} {125147} (\bibinfo {year} {2019})}\BibitemShut {NoStop}%
\bibitem [{\citenamefont {Straquadine}\ \emph {et~al.}(2020)\citenamefont
  {Straquadine}, \citenamefont {Ikeda},\ and\ \citenamefont
  {Fisher}}]{Straquadine2020}%
  \BibitemOpen
  \bibfield  {author} {\bibinfo {author} {\bibfnamefont {J.~A.~W.}\
  \bibnamefont {Straquadine}}, \bibinfo {author} {\bibfnamefont {M.~S.}\
  \bibnamefont {Ikeda}}, \ and\ \bibinfo {author} {\bibfnamefont {I.~R.}\
  \bibnamefont {Fisher}},\ }\href {http://arxiv.org/abs/2005.10461} {\bibfield
  {journal} {\bibinfo  {journal} {arXiv}\ } (\bibinfo {year} {2020})},\ \Eprint
  {http://arxiv.org/abs/2005.10461} {arXiv:2005.10461 [cond-mat]} \BibitemShut
  {NoStop}%
\bibitem [{\citenamefont {Sullivan}\ and\ \citenamefont
  {Seidel}(1968)}]{Sullivan1968}%
  \BibitemOpen
  \bibfield  {author} {\bibinfo {author} {\bibfnamefont {P.~F.}\ \bibnamefont
  {Sullivan}}\ and\ \bibinfo {author} {\bibfnamefont {G.}~\bibnamefont
  {Seidel}},\ }\href@noop {} {\bibfield  {journal} {\bibinfo  {journal} {Phys.
  Rev.}\ }\textbf {\bibinfo {volume} {173}},\ \bibinfo {pages} {679} (\bibinfo
  {year} {1968})}\BibitemShut {NoStop}%
\bibitem [{\citenamefont {Velichkov}(1992)}]{Velichkov1992}%
  \BibitemOpen
  \bibfield  {author} {\bibinfo {author} {\bibfnamefont {I.}~\bibnamefont
  {Velichkov}},\ }\href
  {https://linkinghub.elsevier.com/retrieve/pii/001122759290366I} {\bibfield
  {journal} {\bibinfo  {journal} {Cryogenics}\ }\textbf {\bibinfo {volume}
  {32}},\ \bibinfo {pages} {285} (\bibinfo {year} {1992})}\BibitemShut
  {NoStop}%
\bibitem [{\citenamefont {Aln{\ae}s}\ \emph {et~al.}(2015)\citenamefont
  {Aln{\ae}s}, \citenamefont {Blechta}, \citenamefont {Hake}, \citenamefont
  {Johansson}, \citenamefont {Kehlet}, \citenamefont {Logg}, \citenamefont
  {Richardson}, \citenamefont {Ring}, \citenamefont {Rognes},\ and\
  \citenamefont {Wells}}]{Alnaes2015}%
  \BibitemOpen
  \bibfield  {author} {\bibinfo {author} {\bibfnamefont {M.~S.}\ \bibnamefont
  {Aln{\ae}s}}, \bibinfo {author} {\bibfnamefont {J.}~\bibnamefont {Blechta}},
  \bibinfo {author} {\bibfnamefont {J.}~\bibnamefont {Hake}}, \bibinfo {author}
  {\bibfnamefont {A.}~\bibnamefont {Johansson}}, \bibinfo {author}
  {\bibfnamefont {B.}~\bibnamefont {Kehlet}}, \bibinfo {author} {\bibfnamefont
  {A.}~\bibnamefont {Logg}}, \bibinfo {author} {\bibfnamefont {C.}~\bibnamefont
  {Richardson}}, \bibinfo {author} {\bibfnamefont {J.}~\bibnamefont {Ring}},
  \bibinfo {author} {\bibfnamefont {M.~E.}\ \bibnamefont {Rognes}}, \ and\
  \bibinfo {author} {\bibfnamefont {G.~N.}\ \bibnamefont {Wells}},\ }\href@noop
  {} {\bibfield  {journal} {\bibinfo  {journal} {Arch. Numer. Softw.}\ }\textbf
  {\bibinfo {volume} {3}},\ \bibinfo {pages} {9} (\bibinfo {year}
  {2015})}\BibitemShut {NoStop}%
\bibitem [{\citenamefont {Logg}\ and\ \citenamefont {Wells}(2010)}]{Logg2010}%
  \BibitemOpen
  \bibfield  {author} {\bibinfo {author} {\bibfnamefont {A.}~\bibnamefont
  {Logg}}\ and\ \bibinfo {author} {\bibfnamefont {G.~N.}\ \bibnamefont
  {Wells}},\ }\href {https://dl.acm.org/doi/10.1145/1731022.1731030} {\bibfield
   {journal} {\bibinfo  {journal} {ACM Trans. Math. Softw.}\ }\textbf {\bibinfo
  {volume} {37}},\ \bibinfo {pages} {1} (\bibinfo {year} {2010})}\BibitemShut
  {NoStop}%
\bibitem [{\citenamefont {Geuzaine}\ and\ \citenamefont
  {Remacle}(2009)}]{Remacle2012}%
  \BibitemOpen
  \bibfield  {author} {\bibinfo {author} {\bibfnamefont {C.}~\bibnamefont
  {Geuzaine}}\ and\ \bibinfo {author} {\bibfnamefont {J.}~\bibnamefont
  {Remacle}},\ }\href {http://onlinelibrary.wiley.com/doi/10.1002/nme.3279/full
  http://doi.wiley.com/10.1002/nme.2579} {\bibfield  {journal} {\bibinfo
  {journal} {Int. J. Numer. Methods Eng.}\ }\textbf {\bibinfo {volume} {79}},\
  \bibinfo {pages} {1309} (\bibinfo {year} {2009})}\BibitemShut {NoStop}%
\bibitem [{\citenamefont {Croarkin}\ and\ \citenamefont
  {Guthrie}(1993)}]{Croarkin1993}%
  \BibitemOpen
  \bibfield  {author} {\bibinfo {author} {\bibfnamefont {M.~C.}\ \bibnamefont
  {Croarkin}}\ and\ \bibinfo {author} {\bibfnamefont {W.~F.}\ \bibnamefont
  {Guthrie}},\ }\href {https://srdata.nist.gov/its90/authors/authors.html
  https://srdata.nist.gov/its90/main/its90_main_page.html} {\emph {\bibinfo
  {title} {Natl. Inst. Stand. Technol. Monogr. 175}}},\ \bibinfo {type} {Tech.
  Rep.}\ (\bibinfo  {institution} {National Institute of Standards and
  Technology},\ \bibinfo {address} {Gaithersburg, MD},\ \bibinfo {year}
  {1993})\BibitemShut {NoStop}%
\bibitem [{\citenamefont {Machida}\ \emph {et~al.}(2009)\citenamefont
  {Machida}, \citenamefont {Tomokuni}, \citenamefont {Isono}, \citenamefont
  {Izawa}, \citenamefont {Nakajima},\ and\ \citenamefont
  {Tamegai}}]{Machida2009a}%
  \BibitemOpen
  \bibfield  {author} {\bibinfo {author} {\bibfnamefont {Y.}~\bibnamefont
  {Machida}}, \bibinfo {author} {\bibfnamefont {K.}~\bibnamefont {Tomokuni}},
  \bibinfo {author} {\bibfnamefont {T.}~\bibnamefont {Isono}}, \bibinfo
  {author} {\bibfnamefont {K.}~\bibnamefont {Izawa}}, \bibinfo {author}
  {\bibfnamefont {Y.}~\bibnamefont {Nakajima}}, \ and\ \bibinfo {author}
  {\bibfnamefont {T.}~\bibnamefont {Tamegai}},\ }\href@noop {} {\bibfield
  {journal} {\bibinfo  {journal} {J. Phys. Soc. Japan}\ }\textbf {\bibinfo
  {volume} {78}},\ \bibinfo {pages} {5} (\bibinfo {year} {2009})}\BibitemShut
  {NoStop}%
\bibitem [{\citenamefont {Chu}\ \emph {et~al.}(2009)\citenamefont {Chu},
  \citenamefont {Analytis}, \citenamefont {Kucharczyk},\ and\ \citenamefont
  {Fisher}}]{Chu2009}%
  \BibitemOpen
  \bibfield  {author} {\bibinfo {author} {\bibfnamefont {J.}~\bibnamefont
  {Chu}}, \bibinfo {author} {\bibfnamefont {J.~G.}\ \bibnamefont {Analytis}},
  \bibinfo {author} {\bibfnamefont {C.}~\bibnamefont {Kucharczyk}}, \ and\
  \bibinfo {author} {\bibfnamefont {I.~R.}\ \bibnamefont {Fisher}},\ }\href
  {https://link.aps.org/doi/10.1103/PhysRevB.79.014506} {\bibfield  {journal}
  {\bibinfo  {journal} {Phys. Rev. B}\ }\textbf {\bibinfo {volume} {79}},\
  \bibinfo {pages} {014506} (\bibinfo {year} {2009})}\BibitemShut {NoStop}%
\bibitem [{\citenamefont {Meinero}\ \emph {et~al.}(2019)\citenamefont
  {Meinero}, \citenamefont {Caglieris}, \citenamefont {Pallecchi},
  \citenamefont {Lamura}, \citenamefont {Ishida}, \citenamefont {Eisaki},
  \citenamefont {Continenza},\ and\ \citenamefont {Putti}}]{Meinero2019}%
  \BibitemOpen
  \bibfield  {author} {\bibinfo {author} {\bibfnamefont {M.}~\bibnamefont
  {Meinero}}, \bibinfo {author} {\bibfnamefont {F.}~\bibnamefont {Caglieris}},
  \bibinfo {author} {\bibfnamefont {I.}~\bibnamefont {Pallecchi}}, \bibinfo
  {author} {\bibfnamefont {G.}~\bibnamefont {Lamura}}, \bibinfo {author}
  {\bibfnamefont {S.}~\bibnamefont {Ishida}}, \bibinfo {author} {\bibfnamefont
  {H.}~\bibnamefont {Eisaki}}, \bibinfo {author} {\bibfnamefont
  {A.}~\bibnamefont {Continenza}}, \ and\ \bibinfo {author} {\bibfnamefont
  {M.}~\bibnamefont {Putti}},\ }\href
  {https://iopscience.iop.org/article/10.1088/1361-648X/ab080b} {\bibfield
  {journal} {\bibinfo  {journal} {J. Phys. Condens. Matter}\ }\textbf {\bibinfo
  {volume} {31}},\ \bibinfo {pages} {214003} (\bibinfo {year}
  {2019})}\BibitemShut {NoStop}%
\bibitem [{\citenamefont {Corruccini}\ and\ \citenamefont
  {Gniewek}(1960)}]{Corruccini1960}%
  \BibitemOpen
  \bibfield  {author} {\bibinfo {author} {\bibfnamefont {R.~J.}\ \bibnamefont
  {Corruccini}}\ and\ \bibinfo {author} {\bibfnamefont {J.~J.}\ \bibnamefont
  {Gniewek}},\ }\href
  {https://nvlpubs.nist.gov/nistpubs/Legacy/MONO/nbsmonograph21.pdf} {\emph
  {\bibinfo {title} {NBS Monogr.}}},\ \bibinfo {type} {Tech. Rep.}\ (\bibinfo
  {institution} {National Bureau of Standards},\ \bibinfo {address}
  {Gaithersburg, MD},\ \bibinfo {year} {1960})\BibitemShut {NoStop}%
\bibitem [{\citenamefont {Schwartzberg}(1970)}]{Schwartzberg1970}%
  \BibitemOpen
  \bibfield  {author} {\bibinfo {author} {\bibfnamefont {F.~R.}\ \bibnamefont
  {Schwartzberg}},\ }\href@noop {} {\emph {\bibinfo {title} {Cryogenic
  Materials Data Handbook}}}\ (\bibinfo  {publisher} {National Technical
  Information Service},\ \bibinfo {address} {Springfield, VA},\ \bibinfo {year}
  {1970})\ p.\ \bibinfo {pages} {748}\BibitemShut {NoStop}%
\bibitem [{\citenamefont {Sundqvist}(1992)}]{Sundqvist1992}%
  \BibitemOpen
  \bibfield  {author} {\bibinfo {author} {\bibfnamefont {B.}~\bibnamefont
  {Sundqvist}},\ }\href@noop {} {\bibfield  {journal} {\bibinfo  {journal} {J.
  Appl. Phys.}\ }\textbf {\bibinfo {volume} {72}},\ \bibinfo {pages} {539}
  (\bibinfo {year} {1992})}\BibitemShut {NoStop}%
\bibitem [{\citenamefont {Nakamura}\ \emph {et~al.}(2018)\citenamefont
  {Nakamura}, \citenamefont {Fujii}, \citenamefont {Matsukawa}, \citenamefont
  {Katagiri},\ and\ \citenamefont {Fukuyama}}]{Nakamura2018}%
  \BibitemOpen
  \bibfield  {author} {\bibinfo {author} {\bibfnamefont {S.}~\bibnamefont
  {Nakamura}}, \bibinfo {author} {\bibfnamefont {T.}~\bibnamefont {Fujii}},
  \bibinfo {author} {\bibfnamefont {S.}~\bibnamefont {Matsukawa}}, \bibinfo
  {author} {\bibfnamefont {M.}~\bibnamefont {Katagiri}}, \ and\ \bibinfo
  {author} {\bibfnamefont {H.}~\bibnamefont {Fukuyama}},\ }\href
  {https://doi.org/10.1016/j.cryogenics.2018.09.001
  https://linkinghub.elsevier.com/retrieve/pii/S0011227518300468} {\bibfield
  {journal} {\bibinfo  {journal} {Cryogenics}\ }\textbf {\bibinfo {volume}
  {95}},\ \bibinfo {pages} {76} (\bibinfo {year} {2018})}\BibitemShut {NoStop}%
\bibitem [{\citenamefont {Langtangen}\ and\ \citenamefont
  {Logg}(2016)}]{Langtangen2016}%
  \BibitemOpen
  \bibfield  {author} {\bibinfo {author} {\bibfnamefont {H.~P.}\ \bibnamefont
  {Langtangen}}\ and\ \bibinfo {author} {\bibfnamefont {A.}~\bibnamefont
  {Logg}},\ }\href {http://link.springer.com/10.1007/978-3-319-52462-7} {\emph
  {\bibinfo {title} {Solving {PDEs} in {Python}}}}\ (\bibinfo  {publisher}
  {Springer International Publishing},\ \bibinfo {address} {Cham},\ \bibinfo
  {year} {2016})\BibitemShut {NoStop}%
\bibitem [{\citenamefont {Kopp}\ and\ \citenamefont {Slack}(1971)}]{Kopp1971}%
  \BibitemOpen
  \bibfield  {author} {\bibinfo {author} {\bibfnamefont {J.}~\bibnamefont
  {Kopp}}\ and\ \bibinfo {author} {\bibfnamefont {G.}~\bibnamefont {Slack}},\
  }\href {https://linkinghub.elsevier.com/retrieve/pii/0011227571900051}
  {\bibfield  {journal} {\bibinfo  {journal} {Cryogenics}\ }\textbf {\bibinfo
  {volume} {11}},\ \bibinfo {pages} {22} (\bibinfo {year} {1971})}\BibitemShut
  {NoStop}%
\bibitem [{\citenamefont {Rittel}(1998)}]{Rittel1998}%
  \BibitemOpen
  \bibfield  {author} {\bibinfo {author} {\bibfnamefont {D.}~\bibnamefont
  {Rittel}},\ }\href {http://link.springer.com/10.1007/BF02321647} {\bibfield
  {journal} {\bibinfo  {journal} {Exp. Mech.}\ }\textbf {\bibinfo {volume}
  {38}},\ \bibinfo {pages} {73} (\bibinfo {year} {1998})}\BibitemShut {NoStop}%
\bibitem [{\citenamefont {Henning}\ and\ \citenamefont
  {Parker}(1967)}]{Henning1967}%
  \BibitemOpen
  \bibfield  {author} {\bibinfo {author} {\bibfnamefont {C.~D.}\ \bibnamefont
  {Henning}}\ and\ \bibinfo {author} {\bibfnamefont {R.}~\bibnamefont
  {Parker}},\ }\href
  {https://asmedigitalcollection.asme.org/heattransfer/article/89/2/146/411821/Transient-Response-of-an-Intrinsic-Thermocouple}
  {\bibfield  {journal} {\bibinfo  {journal} {J. Heat Transfer}\ }\textbf
  {\bibinfo {volume} {89}},\ \bibinfo {pages} {146} (\bibinfo {year}
  {1967})}\BibitemShut {NoStop}%
\bibitem [{Note3()}]{Note3}%
  \BibitemOpen
  \bibinfo {note} {The true value is slightly depressed below one, due to the
  finite heat capacity of the thermocouple and bond. Additionally, the boundary
  conditions at the far end of the wire will implement $\Gamma _b(\omega )$
  effects, but at very low frequencies due to the low total
  conductance.}\BibitemShut {Stop}%
\end{thebibliography}%
\end{document}